\documentclass[aps,secnumarabic,nobalancelastpage,nofootinbib]{revtex4}
\usepackage{amsmath,amssymb,amsfonts}
\usepackage{graphicx}
\usepackage{dcolumn}
\usepackage{bm}
\usepackage{hyperref}
\usepackage{bbm}
\usepackage{tikz}
\usepackage{physics}
\usepackage{enumitem}
\usepackage{epsfig}
\usepackage[all]{xy}
\usepackage{amsthm}
\usepackage{dcolumn}
\usepackage{url}
\usepackage{dsfont}
\usepackage{slashed}
\usepackage{mathrsfs}
\usepackage{soul}
\usepackage{float}
\DeclareMathAlphabet{\mathpzc}{OT1}{pzc}{m}{it}

\newcommand{\be}{\begin{equation}}
\newcommand{\ee}{\end{equation}}
\newcommand{\bea}{\begin{eqnarray}}

\newcommand{\eea}{\end{eqnarray}}

\newcommand{\vsi}{\varsigma}

\newcommand{\ds}{\displaystyle}

\def\inbar{\,\vrule height1.5ex width.4pt depth0pt}
\def\IR{\relax{\rm I\kern-.18em R}}
\def\IC{\relax\hbox{$\inbar\kern-.3em{\rm C}$}}

\def\H{\mathbb{H}}
\def\R{\mathbb{R}}
\def\N{\mathbb{N}}
\def\C{\mathbb{C}}
\def\Z{\mathbb{Z}}

\def\ii{\mathrm{i}}
\def\ud{\mathrm{d}}

\def\pz{\pmb{z}}

\def\bpz{\overline{\pmb{z}}}

\def\bu{\mathbbm{1}}

\def\C{{\mathbb C}}

\def\N{{\mathbb N}}
\def\R{{\mathbb R}}

\def\Z{{\mathbb Z}}

\definecolor{hervecolor}{rgb}{0.8,0,0.7}

\begin{document}

\title{Matrix Elements and Characters of the Discrete Series (``Massive'') Unitary Irreducible Representations of Sp$(4,\R)$}

\author{Jean-Pierre Gazeau$^{1}$\footnote{gazeau@apc.in2p3.fr}}

\author{Mariano A. del Olmo$^2$\footnote{marianoantonio.olmo@uva.es}}

\author{Hamed Pejhan$^{3}$\footnote{pejhan@math.bas.bg}}

\affiliation{$^1$Universit\'e Paris Cit\'{e}, CNRS, Astroparticule et Cosmologie, F-75013 Paris, France\\
$^2$Departamento de F\'{\i}sica Te\'orica and IMUVA, Universidad de Valladolid, E-47011, Valladolid, Spain\\
$^3$Institute of Mathematics and Informatics, Bulgarian Academy of Sciences, Acad. G. Bonchev Str. Bl. 8, 1113, Sofia, Bulgaria}

\date{\today}

\begin{abstract}
This paper obtains the matrix elements and characters of the discrete series unitary irreducible representations (UIRs) of the Sp$(4,\mathbb{R})$ group. With an isomorphic relationship to the two-fold covering of SO$_0(2,3)$ (Sp$(4,\mathbb{R}) \sim$ SO$_0(2,3)\times \mathbb{Z}_2$), this group holds particular importance as the kinematical/relativity group within the framework of ($1+3$-dimensional) anti-de Sitter spacetime.
\end{abstract}

\maketitle

\setcounter{equation}{0}
 \section{Introduction}
In a recent paper \cite{AdS00}, we delved into the relativity group of $(1+3)$-dimensional anti-de Sitter spacetime. Specifically, our examination focused on the Sp$(4,\mathbb{R})$ group, which shares an isomorphism with the two-fold covering of SO$_0(2,3)$, serving as the kinematical/relativity group for anti-de Sitter spacetime. Our primary emphasis was on the geometry of the domain $\mathcal{D}^{(3)}
 \sim \mathrm{Sp}(4,\mathbb{R})/\mathrm{S}(\mathrm{U}(2)\times\mathrm{U}(1))$, which is  a Sp$(4,\mathbb{R})$ left coset derived from the Cartan decomposition of the group, with the group's customary action on it.
 We underscored the K\"{a}hlerian structure and the phase-space interpretation of $\mathcal{D}^{(3)}$, which governs the classical phase space for the set of free motions of a test massive particle on anti-de Sitter spacetime. 

Moreover, we introduced the discrete series (in a broad sense) of Sp$(4,\mathbb{R})$ representations, which operate on the (Segal-)Bargmann-Fock Hilbert spaces of holomorphic functions within the Cartan domain $\mathcal{D}^{(3)}$. 

Our discussion further explored the Poincar\'{e} and the Newton-Hooke contraction limits \cite{inonu52,bacryjmll68,deromedubois,dubois} of the construction, spanning both classical and quantum levels. Of particular interest is the Poincar\'{e} contraction limit, which gives rise to the phase space for massive elementary systems under the Poincar\'{e} kinematical symmetry, represented by the well-known mass shell hyperboloid equation $E^2 - (\vec{p}\cdot\vec{p})c^2 = m^2c^4$, at the classical level. On the quantum level, the representations contract the massive Poincar\'{e} UIRs with positive mass and positive energy in such a way that they exhaust the entire set of the latter. Thus, we term the corresponding discrete series UIRs the ``massive'' representations of Sp$(4,\mathbb{R})$.

In this supplement paper, we precisely obtain the matrix elements and the characters of these Sp$(4,\mathbb{R})$ ``massive'' UIRs.
These two companion papers are intended to serve as the primary steps towards implementing the covariant integral quantization of anti-de Sitterian ``massive'' elementary systems, specifically, the implementation of the Sp$(4,\mathbb{R})$ covariant integral quantization of functions on the Cartan domain $\mathcal{D}^{(3)}$. The main goal is to draw a parallel to a previous article \cite{olmogaz19} by del Olmo and Gazeau, which was dedicated to such a study in the $(1+1)$-dimensional anti-de Sitter spacetime. In a broader context, however, these two papers are components of a series of books/papers that contribute to a broader research plan (see Ref. \cite{Gazeau2022} and references therein) aimed at developing a consistent formulation, on both classical and quantum levels, of elementary systems in the global structure of de Sitter and anti-de Sitter spacetimes.

To accomplish our objective, the remainder of this paper is structured as follows. Firstly, in section \ref{Sec. 2}, to ensure the self-sufficiency of this study, we provide a brief review of all the necessary mathematical concepts drawn from our previous article \cite{AdS00}. Sections \ref{Sec. M elements} and \ref{Sec. characters} are dedicated, respectively, to a detailed computation of the matrix elements and characters of the UIRs. Finally, in section \ref{Sec.:Summary}, we offer a concise summary of our findings. We also present three appendices to make more complete the paper. In Appendix \ref{App. quaternions}, we present a brief introduction to the complex quaternions and in Appendix 
\ref{App. SU(2)} the matrix elements of the  UIRs of SU$(2)$ as well as of their tensor product considering the elements of  SU$(2)$ as real quaternions. Hereafter, we make the  holomorphic extension to the complex quaternions. Finally, Appendix \ref{App. expansion} is devoted to obtain an expansion formula for the determinant of the complex quaternionic  $(1+z\,\underline z)^{-\lambda}$ (with $\lambda\neq 0$ a positive integer) using the generating function for Gegenbauer polynomials. 

\setcounter{equation}{0} 
\section{The Sp$(4,\R)$ group and its representation in the discrete series} \label{Sec. 2}

The materials presented in this section have been sourced from the preceding companion paper \cite{AdS00}.

With respect to the complex quaternionic algebra introduced in appendix \ref{App. quaternions}, the elements of the Sp$(4,\R)$ group are represented by $2 \times 2$-complex quaternionic matrices of the following form:
\begin{equation}\label{sp4Rdef}
g =\begin{pmatrix} a & b \\ -\overline b & \overline a \end{pmatrix}, \quad a, b \in \H_{\C}\,, \qquad 
\mathsf{e} \equiv \begin{pmatrix} 1 & 0 \\  0 &  1 \end{pmatrix},
\end{equation}
where $\overline z$ means the complex conjugate of $z$ (see \eqref{conjugate}) and $\mathsf{e}$ is the identity element.
The inverse $g^{-1}$ is obtain by:
\begin{equation}\label{sp4inv}
g^{-1} = \begin{pmatrix} 0 & 1 \\ 1 & 0 \end{pmatrix}\,{}^{\mathrm{t}}\widetilde{g}\,
\begin{pmatrix} 0 & 1 \\ 1 & 0 \end{pmatrix} =
\begin{pmatrix} a^{\ast} & \widetilde b \\ -b^{\ast} & \widetilde a \end{pmatrix},
\end{equation}
where $\widetilde{g}$ stands for the quaternionic conjugate of $g$ \eqref{conjugate} and  ${}^{\mathrm{t}}\widetilde{g}$ for the transpose of $\widetilde g$. Then, since  $g\,g^{-1}=\mathsf{e}$ or  $g^{-1}g=\mathsf{e}$, the complex quaternionic entries have to obey, respectively,
\begin{eqnarray}\label{ggm1}
aa^{\ast}-bb^{\ast} &=& 1\,, \hskip1.15cm a\widetilde b = -b\widetilde a\,,\\[0.2cm]
a^{\ast}a-\widetilde b\, \overline b &=& 1\,,  \hskip1cm  a^{\ast} b = -\widetilde b\overline a\,. \label{ggm2}
\end{eqnarray}

Given the above considerations, it is straightforward to demonstrate that the generic element \eqref{sp4Rdef} of the Sp$(4,\mathbb{R})$ group has a determinant of $1$ (see Ref. \cite{AdS00}).

As already pointed out, this paper \cite{AdS00} particularly deals with the \emph{``massive"}\footnote{In the sense of the Poincar\'e contraction limit \cite{AdS00}.} Sp$(4,\R)$ UIRs which lie in the discrete series representations of this group, more accurately, of its universal covering group $\overline{\mathrm{Sp}(4,\R)}$ \cite{evans67, fronsdal74, baelgagi92, Gazeau2022}. These ``massive" UIRs are labelled by the dimensionless parameters $\varsigma \in \R_+$ and $s\in \N/2$ (spin), while $\varsigma > s +1$. For $\varsigma > s +2$, the representation Hilbert spaces are characterized by (Segal-)Bargmann-Fock spaces $\mathcal{F}^{(\varsigma,s)}$. The latter are the spaces of holomorphic $(2s+1)$-vector functions:
\begin{equation} \label{FBUIR}
\mathcal{D}^{(3)} \ni \pmb{z} \;\mapsto\; f(\pmb{z}) \in \C^{2s+1}\,,
\end{equation}
which are square integrable according to the bilinear form:
\begin{equation} \label{FBinner}
(f_1,f_2)_{(\varsigma,s)} = \mathcal{N}(\varsigma,s) \int_{\mathcal{D}^{(3)}} f_1(\pz)^{\dag}\,D^s\left(\frac{1}{1+\pz\bpz}\right)\,f_2(\pz)\,\big[\det(1+\pz\bpz)\big]^{\varsigma +s-3}\,\dot{\pz} \,, \quad \dot{\pz} \equiv \ud^3 \pmb{x}\, \ud^3 \pmb{y}\,.
\end{equation}
Note that:

 (i) $\mathcal{D}^{(3)}$ denotes the bounded domain of $\C^3$ defined by:
\begin{equation}\label{domC3}
\mathcal{D}^{(3)}= \Big\{ \pz\,;\, \bu_2 - \pmb{Z}\pmb{Z}^{\dag} >0 \Big\} \sim \mathrm{Sp}(4,\R)/\mathrm{S}\big(\mathrm{U}(1)\times \mathrm{SU}(2)\big)\,,
\end{equation}
where we remind the one-to-one map $\pz \mapsto \pmb{Z}(\pz)= \begin{pmatrix} \ii z^3 & \ii z^1 -z^2 \\\ii z^1+z^2 &  - \ii z^3 \end{pmatrix} \equiv \pmb{Z}$ (see Appendix \ref{App. quaternions}). 

(ii) $D^s$ denotes the operator of the irreducible $(2s+1)\times (2s+1)$-matrix representation of the SU$(2)$ group, the holomorphically extended to $\mathrm{M}(2,\C)$ (see Appendix \ref{App. SU(2)}). 

(iii) The normalization constant $\mathcal{N}(\varsigma,s)$ is \cite{onofri76}:
\begin{equation} \label{Nvzs}
\mathcal{N}(\varsigma,s) = \frac{8}{\pi^3}\,\left(\varsigma + s -\frac{3}{2}\right)\,(\varsigma-s-1)\,(\varsigma-s-2)\,.
\end{equation}

(iv) $e_{s\rho}$, with $-s\leqslant \rho \leqslant s$, is given by:
\begin{equation}\begin{array}{lll} \label{essig}
&\mbox{for integer $s$}:\quad\qquad & e_{s\rho} \equiv \left(\delta_{s,\rho},\, \delta_{s-1,\rho} ,\,   \dots\, ,\, \delta_{1,\rho} ,\, \delta_{0,\rho} ,\, \delta_{-1,\rho},\, \dots\, ,\, \delta_{-s+1,\rho} ,\, \delta_{-s,\rho} \right)^{\mathrm{t}}\,,\\[0.3cm]
&\mbox{for half-integer $s$}:\quad\qquad & e_{s\rho} \equiv \left( \delta_{s,\rho},\, \delta_{s-1,\rho} ,\,\dots \,,\, \delta_{\frac{1}{2},\rho} ,\,\delta_{-\frac{1}{2},\rho},\, \dots\, ,\,\delta_{-s+1,\rho} ,\, \delta_{-s,\rho} \right)^{\mathrm{t}}.
\end{array}\end{equation}

(v) The inner product should be modified when $s+1\leqslant\varsigma \leqslant s +2$; here, the Shilov boundary $[0,\pi]\times \mathbb{S}^2$ (namely, the Lie sphere) of $\mathcal{D}^{(3)}$ has a significant r\^ole \cite{onofri76}.

Here, one must notice that the (Segal-)Bargmann-Fock spaces $\mathcal{F}^{(\vsi,s)}$ are actually reproducing-kernel spaces, with the $(2s+1)\times (2s+1)$-matrix-valued reproducing kernel:
\begin{equation} \label{repkerFB}
K^{(\vsi,s)}(\pz,\overline{\pz^{\prime}}) = \big[\det(1+\pz\overline{\pz^{\prime}})\big]^{-\vsi -s}\,D^s\left(1+\pz\overline{\pz^{\prime}}\right)\,,
\end{equation}
while the reproducing property is:
\begin{equation} \label{repprop}
f(\pz) = \left({K^{(\vsi,s)}(\pz,\cdot)}^{\dag},f\right)_{(\vsi,s)}\,, \quad \forall\,f\in \mathcal{F}^{(\vsi,s)}\,.
\end{equation}
The separating expansion of the reproducing kernel \eqref{repkerFB} as below gives us an orthonormal basis for $\mathcal{F}^{(\vsi,s)}$:
\begin{equation} \label{orthobasisFB}
K^{(\vsi,s)}(\pz,\pz^{\prime}) = \sum_\nu \mathrm{F}^{(\vsi,s)}_{\nu}(\pz)\, {\mathrm{F}^{(\vsi,s)}_{\nu}(\pz^{\prime})}^{\dag}\,,
\end{equation}
where the $(2s+1)$-vector-valued analytic $\mathrm{F}^{(\vsi,s)}_{\nu}$s are:
\begin{eqnarray}\label{Enus0}
\mathrm{F}^{(\vsi,s)}_{\nu}(\pz) \equiv \mathrm{F}^{(\vsi,s)}_{l,k,J,M}(\pz) = \sqrt{a_{\varsigma+s,l,k}}\, (\pz\cdot\pz)^{k}\, \mathcal{Y}_{s,l-2k,JM}(\pz) \,,
\end{eqnarray}
while, for a given $s$, we have:
\begin{align}
\mathcal{Y}_{s,l-2k,JM}(\pz) &= \sum_{m\,\rho} (s\rho ,l-2k, m| s ,l-2k, J M)\,e_{s\rho}\, \mathcal{Y}_{l-2k,m}(\pz)\,,\\
\label{converse} \mathcal{Y}_{l-2k,m}(\pz) &= \sum_{JM} \overline{(s\rho, l-2k, m| s, l-2k, J M)}\, e_{s\rho}^\dagger\, \mathcal{Y}_{s,l-2k,JM}(\pz)\,,
\end{align}
where:
\begin{align}
\label{nkm1m2} & l\in \N\,, \ 0\leqslant k\leqslant \left\lfloor \frac{l}{2}\right\rfloor\,, \ |l-2k-s| \leqslant J \leqslant l-2k+s\,, \ M=m+\rho\,, \ \mbox{with} \ 2k-l\leqslant m \leqslant l-2k\,, \ -s\leqslant \rho \leqslant s\,, \ \\
\label{dddddd} &a_{\varsigma+s,l,k} = \frac{2^{2(\varsigma+s)-1}\pi \,\Gamma(k+\varsigma+s - 1/2)\, \Gamma(\varsigma+s + l - k)}{\Gamma(2(\varsigma+s)-1)\, k!\, \Gamma(l-k + 3/2)}\,,
\end{align}
and $(s\rho, l-2k, m| s, l-2k, J M)$ denotes the Clebsch-Gordan or Wigner coefficients (see Eq. \eqref{3japc}) and $\mathcal{Y}_{l-2k,m}(\pz)$ represents the holomorphic extensions to $\C^3$ of the solid spherical harmonics \cite{hassan80}  multiplied by even powers of the Euclidean distance in $\R^3$:
\begin{equation}\label{solspharmp}
\begin{split}
\mathcal{Y}_{lm}(\pz) = (-1)^m\left[\frac{(2l+1)(l-m)!}{4\pi(l+m)!}\right]^{1/2}\, 2^l(z^1+\ii z^2)^m\, \sum_{k=m}^l \frac{\Gamma\left(\frac{k+l+1}{2}\right)}{\Gamma\left(\frac{k-l+1}{2}\right)}\frac{(z^3)^{k-m}}{(k-m)!}\,\frac{(\pz\cdot\pz)^{\frac{l-k}{2}}}{(l-k)!}\,,
\end{split}
\end{equation}
with $\pz = (z^1,z^2,z^3) \in \mathcal{D}^{(3)}$.
Moreover, it is worth noting that:
\begin{equation}\label{HYHY3j}
\begin{array}{lll}
\mathcal{Y}_{l_1m_1}(\pz)\,\mathcal{Y}_{l_2m_2}(\pz)& =&\ds (-1)^{m_3}\sum_{l_3}\left[\frac{(2l_1+1)(2l_2+1)(2l_3+1)}{4\pi}\right]^{1/2} \\[0.3cm] 
&&\quad\ds (\pz\cdot\pz)^{\frac{l_1 +l_2 -l_3}{2}} \begin{pmatrix} l_1 & l_2 & l_3 \\ m_1 & m_2 & -m_3 \end{pmatrix}\,\begin{pmatrix} l_1 & l_2 & l_3 \\ 0 & 0 & 0 \end{pmatrix}\,\mathcal{Y}_{l_3m_3}(\pz)\,.
\end{array}
\end{equation}
where $l_1 +l_2 -l_3$ has to be even.

Here, it is perhaps worth noting that, for the specific case of the scalar ($s=0$) UIRs, the analytic functions $\mathrm{F}^{(\vsi,s=0)}_{\nu}$ simply reduce to:
\begin{align}\label{havich}
\mathrm{F}^{(\varsigma,0)}_{\nu}(\pz) \equiv \mathrm{F}^{(\varsigma,0)}_{l,k,m}(\pz) = \sqrt{a_{\varsigma,l,k}}\, (\pz\cdot\pz)^{k}\, \mathcal{Y}_{l-2k,m}(\pz)\,.
\end{align}
In this context, the $\mathrm{Sp}(4,\R)$-representation operators $U^{(\vsi,s)}(g)$ acting on the support spaces $\mathcal{F}^{(\vsi,s)}$ read as:
\begin{equation} \label{UIRAdS}
f^{\prime} (\pz) = \left(U^{(\vsi,s)}(g)\,f\right) (\pz) = \big[\det(-\overline{b}\pz + \overline{a})\big]^{-\vsi -s}\,D^s(\pz b^{\ast} + a^{\ast})\,f\left(g^{-1}\diamond \pz\right)\,,
\end{equation}
with $g^{-1} = \begin{pmatrix} a & b \\ - \overline{b} & \overline{a} \end{pmatrix}$ and $g^{-1}\diamond\pz = (a\pz +b)(-\overline b \pz + \overline a)^{-1}$.

\setcounter{equation}{0} 
\section{Matrix elements of the Sp$(4,\R)$ representations $U^{(\varsigma,s)}$} \label{Sec. M elements}


\subsection{Scalar representations $U^{(\varsigma,0)}$}

We begin with the Sp$(4,\R)$ scalar UIR $U^{(\varsigma,0)}$, for which, according to Eq.~\eqref{UIRAdS}, we have:
\begin{equation}
\label{12}
f^{\prime} (\pz) = \left(U^{(\vsi,0)}(g)\,f\right) (\pz) = \big[\det(-\overline{b}\pz + \overline{a})\big]^{-\vsi}\, f\left(g^{-1}\diamond \pz\right)\,,
\qquad  f\in \mathcal{F}^{(\vsi,0)}\,,
\end{equation}
with now  $\varsigma > 2$. On the other hand, the corresponding orthonormal basis of the (Segal-)Bargmann-Fock Hilbert space $\mathcal{F}^{(\vsi,0)}$ is given in terms of the set of holomorphic functions $\mathrm{F}^{(\varsigma,0)}_{\nu}(\pz)$ (see Eq. \eqref{havich}). With respect to this basis, we here calculate the matrix elements of the (unitary or not) scalar representation $U^{(\varsigma,0)}$, which requires the expansion of the following formula (issued from the right-hand side of \eqref{12}):
\begin{align} \label{matelc10}
&\sqrt{a_{\varsigma,l,k}}\,\, \big[\det(-\overline{b}\pz + \overline{a})\big]^{-\varsigma}\, \,\;\; {\big((g^{-1}\diamond \pz)\cdot (g^{-1}\diamond \pz)\big)^{k}\,} \,\;\; \mathcal{Y}_{l-2k,m}\big(g^{-1}\diamond \pz\big) \nonumber\\[0.2cm]
&\quad\quad = \sqrt{a_{\varsigma,l,k}}\,\, \big[\det(-\overline{b}\pz + \overline{a})\big]^{-\varsigma} \,\;\; \big((g^{-1}\diamond \pz) \widetilde{(g^{-1}\diamond \pz)}\big)^{k} \,\;\; \mathcal{Y}_{l-2k,m}\big(g^{-1}\diamond \pz\big) \nonumber\\[0.2cm]
&\quad\quad = \sqrt{a_{\varsigma,l,k}}\,\big[\det(-\overline{b}\pz + \overline{a})\big]^{-\varsigma} \,\;\; \big[ \det(g^{-1}\diamond \pz) \big]^{k}\,\;\; \mathcal{Y}_{l-2k,m}\big(g^{-1}\diamond \pz\big) \nonumber\\[0.2cm]
&\quad\quad = \sqrt{a_{\varsigma,l,k}}\,\, \big[\det(-\overline{b}\pz + \overline{a})\big]^{-\varsigma} \,\;\; \big[ \det(a\pz +b)\big]^k\, \big[\det(-\overline b \pz + \overline a)\big]^{-k} \,\;\; \mathcal{Y}_{l-2k,m}\big((a\pz +b)(-\overline b \pz + \overline a)^{-1}\big) \,, \nonumber\\[0.2cm]
&\quad\quad = \sqrt{a_{\varsigma,l,k}}\,\, \underbrace{\big[\det(-\overline{b}\pz + \overline{a})\big]^{-\varsigma-k+1}}_{\equiv{\cal{A}}} \, \underbrace{\big[ \det(a\pz +b)\big]^k\,}_{\equiv{\cal{B}}}\, \underbrace{\big[\det(-\overline{b}\pz + \overline{a})\big]^{-1} \, \mathcal{Y}_{l-2k,m}\big((a\pz +b)(-\overline b \pz + \overline a)^{-1}\big)}_{\equiv{\cal{C}}} \,, 
\end{align}
where, to get the second line from the first one, we have used the identity \eqref{zzp} along with the fact that the term $g^{-1}\diamond \pz$ is a pure-vector (complex) quaternion (i.e., $z= \underbrace{(0,\pmb{z})}\equiv \pmb{z}$). 

We now look at the different terms of \eqref{matelc10}:
\medskip

\textbf{\emph{Term ${\cal{A}}$.}} We invoke the expansion formula \eqref{expdetlambd'}. Actually, for an invertible $\overline{a}$ (which is always the case for $g\in$Sp$(4,\R)$), we have:
\begin{eqnarray}
{\cal{A}} \equiv \big[\det(-\overline{b}\pz + \overline{a})\big]^{-\varsigma-k +1} &=& \big[\det\overline{a}\big]^{-\varsigma-k +1} \; \big[\det(1 - {\overline{a}}^{-1}\overline{b}\pz )\big]^{-\varsigma-k +1} \nonumber\\
&=& \big[\det\overline{a}\big]^{-\varsigma-k +1} \; \big[\det(1 + \pz\, \overline{\textbf{w}})\big]^{-\varsigma-k +1}\,, \qquad \textbf{w} \equiv \widetilde b {\widetilde a}^{-1}\,.
\end{eqnarray}
Note that, to get the second line from the first one, we have used the fact that $\det z = \det \widetilde{z}$, for all $z\in\H_{\C}$. Moreover, one should notice that $\textbf{w}$ is a pure-vector (complex) quaternion; to see the point, we refer to the second identity in \eqref{ggm1},\footnote{According to the second identity in \eqref{ggm1}, we have $a\widetilde b = - b\widetilde a$. Now, by multiplying the latter from the left by $a^{-1}$ and subsequently from the right by ${\widetilde a}^{-1}$, we get $\widetilde b {\widetilde a}^{-1} = - a^{-1} b$.} based upon which we have $\widetilde{(\widetilde b {\widetilde a}^{-1})} = \widetilde{(-a^{-1}b)} = - \widetilde b {\widetilde a}^{-1}$. It is also worth mentioning that, the pure-vector (complex) quaternion $\textbf{w}$ lies in the domain ${\cal{D}}^{(3)}$; this point can be easily understood from the first identity in \eqref{ggm1}. From the expansion formula \eqref{expdetlambd'}, we then get:
\begin{equation}\label{expdetlambd''}
{\cal{A}} = 
\big[\det\overline{a}\big]^{-\varsigma-k +1} \; 
\sum_{\mathfrak{l}=0}^{\infty}
\sum_{\mathfrak{k} = 0}^{\lfloor \frac{\mathfrak{l}}{2}\rfloor} 
\sum_{\mathfrak{m}=2\mathfrak{k}-\mathfrak{l}}^{\mathfrak{l}-2\mathfrak{k}} \,a_{(\varsigma,k),\mathfrak{l},\mathfrak{k}}\,(\pz\cdot\pz)^{\mathfrak{k}}\, \mathcal{Y}_{\mathfrak{l}-2\mathfrak{k},\mathfrak{m}}(\pz)\,\overline{(\textbf{w}\cdot\textbf{w})}^{\mathfrak{k}}\, \overline{\mathcal{Y}_{\mathfrak{l}-2\mathfrak{k},\mathfrak{m}}(\textbf{w})}\,,\quad
\end{equation}
with:
\begin{eqnarray} \label{normlk D}
a^{}_{(\varsigma,k),\mathfrak{l},\mathfrak{k}} = \frac{2^{2\varsigma+2k-3} \;\pi \,\Gamma\left(\mathfrak{k}+\varsigma+k - \frac{3}{2}\right) \Gamma(\varsigma+k + \mathfrak{l} - \mathfrak{k}-1)}{\Gamma(2\varsigma+2k-3)\,\mathfrak{k}! \,\Gamma\left(\mathfrak{l}-\mathfrak{k} + \frac{3}{2}\right)}\,.
\end{eqnarray}

\textbf{\emph{Term ${\cal{B}}$.}} The expansion of term ${\cal{B}} \equiv \big[ \det(a\pz +b)\big]^k$ requires sketching steps parallel to what we have done above:\footnote{Note that $b \; (\neq 0)$ is invertible for any $g\in$ Sp$(4,\R)$; the cases involving $b=0$ must be treated individually from the very beginning, as we will discuss in the sequel.}
\begin{eqnarray}\label{ewew}
{\cal{B}} \equiv \big[ \det(a\pz +b)\big]^k &=& \big[\det b \big]^k \, \big[ \det(1+b^{-1}a\pz) \big]^k \nonumber\\
&=& \big[\det b \big]^k \, \big[ \det(1+ \pz \, \overline{\textbf{u}}) \big]^k\,, \qquad \textbf{u} \equiv \overline{b}^{-1}\overline{a}\,.
\end{eqnarray}
Again, to get the second line from the first one, we have used the fact that $\det z = \det \widetilde{z}$, for all $z\in\H_{\C}$, while we have noticed that, as a by-product of the second identity in \eqref{ggm2}, we have $b^\ast \overline a = - a^\ast \overline{b}$, and therefore, $\widetilde{(b^\ast \overline a)}= \widetilde{(- a^\ast \overline{b})} = - b^\ast \overline a$, which means that, $b^\ast \overline a$ and consequently $\overline{b}^{-1}\overline{a} \; (\equiv \textbf{u})$ are pure-vector (complex) quaternions. Then, taking parallel steps to those that have led to Eq.~\eqref{expdetlambd'}, but this time with respect to Eqs.~\eqref{234} (while we have in mind Eq. \eqref{footnote negative lambda0}), we can expand the above formula as:
\begin{align}
{\cal{B}} \equiv \big[ \det(a\pz +b)\big]^k = \big[\det b \big]^k \, \sum_{\mathfrak{\underline{l}}=0}^{2k}\sum_{\mathfrak{\underline{k}} = 0}^{\lfloor \frac{\mathfrak{\underline{l}}}{2}\rfloor} \sum_{\mathfrak{\underline{m}}=2\mathfrak{\underline{k}}-\mathfrak{\underline{l}}}^{\mathfrak{\underline{l}}-2\mathfrak{\underline{k}}} \,a^\prime_{k,\mathfrak{\underline{l}},\mathfrak{\underline{k}}}\,(\pz\cdot\pz)^{\mathfrak{\underline{k}}}\, \mathcal{Y}_{\mathfrak{\underline{l}}-2\mathfrak{\underline{k}},\mathfrak{\underline{m}}}(\pz)\, \overline{(\textbf{u}\cdot\textbf{u})}^{\mathfrak{\underline{k}}}\, \overline{\mathcal{Y}_{\mathfrak{\underline{l}}-2\mathfrak{\underline{k}},\mathfrak{\underline{m}}}(\textbf{u})}\,,
\end{align}
with:
\begin{equation} \label{normlk D1}
a^{\prime}_{k,\mathfrak{\underline{l}},\mathfrak{\underline{k}}} =
 \frac{2\pi^{3/2}\,(-k-1/2)_{\mathfrak{\underline{k}}}\,(-k)_{\mathfrak{\underline{l}}-\mathfrak{\underline{k}}}}{\mathfrak{\underline{k}}! \,\Gamma\left(\mathfrak{\underline{l}}-\mathfrak{\underline{k}} + \frac{3}{2}\right)}\,
\end{equation}
where  $(x)_k$ is the Pochhammer symbol \eqref{pochhammer}. Here, we would like to point out that the above extension formula, contrary to its twin \eqref{expdetlambd''}, is a finite series.
\medskip

\textbf{\emph{Term ${\cal{C}}$.}} Employing Eq. \eqref{YlDl2}, we first rewrite term ${\cal{C}}$ in the following form:
\begin{eqnarray}\label{131313}
{\cal{C}} &\equiv& \big[\det(-\overline{b}\pz + \overline{a})\big]^{-1} \, \mathcal{Y}_{l-2k,m}\big((a\pz +b)(-\overline b \pz + \overline a)^{-1}\big) \nonumber\\
&=& \sum_{\stackrel{m_1 m_2}{m_2-m_1=m}} A_{l,k,m^{}_1,m^{}_2} \, \big[\det(-\overline{b}\pz + \overline{a})\big]^{-1} \, D_{m_1m_2}^{l/2-k}\big[\big(z^4,(a\pz +b)(-\overline b \pz + \overline a)^{-1}\big)\big]\,,
\end{eqnarray}
where $z^4$ is arbitrary, and hence, we shall set it to zero,\footnote{Note that this choice allows going back to $\mathcal{Y}_{l-2k,m}$ through the relation \eqref{DlYl2}.} and:
\begin{align} \label{normlk D2}
A_{l,k,m_1,m_2} = 2^{-l+2k}\, \sigma^{l-2k}_{m=m_2-m_1} \, \left(\frac{2l-4k+1}{4\pi}\right)^{1/2} (-1)^{m_1} \; 
\big(\sigma^{l/2-k}_{m_1 ,m_2}\big)^{-1} \,,
\end{align}
where $\sigma^{l}_{m}$ and $\sigma^{l}_{m_1 ,m_2}$  are given by \eqref{sjm1}.
Note that below, by abuse of notations, we will denote $D_{m_1m_2}^{l/2-k}\big[\big(0,(a\pz +b)(-\overline b \pz + \overline a)^{-1}\big)\big]$ by $D_{m_1m_2}^{l/2-k}\big((a\pz +b)(-\overline b \pz + \overline a)^{-1}\big)$. Now, invoking the relations \eqref{addexp1}, \eqref{addexp2}, and \eqref{multexp}, while $L/2 \equiv {l}/{2}-k$, we get:
\begin{eqnarray} \label{matelc1}
\mbox{Eq. \eqref{131313}} &=& \sum_{\stackrel{m_1 m_2}{m_2-m_1=m}} A_{l,k,m^{}_1,m^{}_2} \, 
\frac{\sigma_{m_2}^{L/2}}{\sigma_{m_1}^{L/2}} \;\; 
\sum_{\stackrel{L_i m_{i1} m_{i2}}{i=1,\dots,4}}
\delta^{}_{L,L_1+L_2}\; \delta^{}_{L_4,L+L_3}\; \delta^{}_{m_{41},m_{32}+m_{22}+m_{12}}\; \delta^{}_{m_{42},m_2+m_{31}}\; \delta^{}_{m_{1},m_{11}+m_{21}} \nonumber\\
&&\qquad\qquad \times \; \sigma^{L_1/2}_{m_{11},m_{12}} \; D^{L_1/2}_{m_{11}m_{12}}(a\pz) \; \sigma^{L_2/2}_{m_{21},m_{22}} \; D^{L_2/2}_{m_{21}m_{22}}(b) \nonumber\\
&& \qquad\qquad\times \; (-1)^{L_3} \; \sigma^{L_3/2}_{m_{31},m_{32}} \; D^{L_3/2}_{m_{31}m_{32}}(-\overline{b}\pz)\; \Big(\sigma^{L_4/2}_{m_{41},m_{42}} \Big)^{-1} \; (\det \overline{a})^{-1} \; D^{L_4/2}_{m_{41}m_{42}}({\overline{a}}^{-1}) \,.\;\;\;\;\;\;\;\;\;\;
\end{eqnarray}
Finally, respectively using Eqs. \eqref{multexp}, \eqref{DlYl2}, and \eqref{HYHY3j}, we obtain:
\begin{align} \label{matelc1'''}
{\cal{C}} 
= \sum_{\stackrel{L^\prime L^{\prime\prime}m^{\prime\prime}}{L^\prime-L^{\prime\prime} = \mathrm{even}}} U_{L, k, m ; L^\prime, L^{\prime\prime}, m^{\prime\prime}} (g) \;\; (\pz\cdot\pz)^{\frac{L^\prime-L^{\prime\prime}}{2}}\mathcal{Y}_{L^{\prime\prime}m^{\prime\prime}}(\pmb{z})\,,
\end{align}
with:
\begin{eqnarray}
U_{L, k, m ; L^\prime, L^{\prime\prime}, m^{\prime\prime}} (g) &=& \sum_{\stackrel{m_1 m_2}{m_2-m_1=m}} A_{l,k,m^{}_1,m^{}_2} \, \frac{\sigma_{m_2}^{L/2}}{\sigma_{m_1}^{L/2}} \; \sum_{\stackrel{L_i m_{i1} m_{i2}}{i=1,2,\dots,4}} \sum_{n\,n^\prime} \delta^{}_{L,L_1+L_2}\; \delta^{}_{L_4,L+L_3}\; \delta^{}_{m_{41},m_{32}+m_{22}+m_{12}}\; \delta^{}_{m_{42},m_2+m_{31}} \nonumber\\
&& \qquad\times \; \delta^{}_{m_{1},m_{11}+m_{21}} \; \delta^{}_{L^\prime,L_1+L_3} \; \sigma^{L_1/2}_{m_{11},m_{12}} \; D^{L_1/2}_{m_{11}n}(a) \; \sigma^{L_2/2}_{m_{21},m_{22}} \; D^{L_2/2}_{m_{21}m_{22}}(b) \nonumber\\
&& \qquad\times \; (-1)^{L_3} \; \sigma^{L_3/2}_{m_{31},m_{32}} \; D^{L_3/2}_{m_{31}n^\prime}(-\overline{b})\; \Big( \sigma^{L_4/2}_{m_{41},m_{42}} \Big)^{-1} \; (\det \overline{a})^{-1} \; D^{L_4/2}_{m_{41}m_{42}}({\overline{a}}^{-1}) \nonumber\\
&&\qquad \times \; \sum_{\stackrel{L^\prime_1}{L^{}_1-L^\prime_1=\mathrm{even}}} (4\pi)^{1/2}\, (-1)^{L_1^{\prime}+m^{}_{12}}\; 2^{L^{\prime}_1}\; (2L^{\prime}_1+1)^{1/2} \left(\frac{(L^{}_1-L^{\prime}_1)!}{(L^{}_1+L^{\prime}_1+1)!}\right)^{1/2} \nonumber\\
&&\quad \times \; \frac{\left[\frac{1}{2}(L^{}_1+L^{\prime}_1)\right]!}{\left[\frac{1}{2}(L^{}_1-L^{\prime}_1)\right]!} \, \begin{pmatrix} {L^{}_1}/{2} & {L^{}_1}/{2}& L^{\prime}_1 \\ n & -m^{}_{12} & m^{\prime}_1 \end{pmatrix} \; \sum_{\stackrel{L^\prime_3}{L^{}_3-L^\prime_3=\mathrm{even}}} (4\pi)^{1/2}\, (-1)^{L_3^{\prime}+m^{}_{32}}\; 2^{L^{\prime}_3}\; (2L^{\prime}_3+1)^{1/2} \nonumber\\
&&\qquad \times \; \left(\frac{(L^{}_3-L^{\prime}_3)!}{(L^{}_3+L^{\prime}_3+1)!}\right)^{1/2} \, \frac{\left[\frac{1}{2}(L^{}_3+L^{\prime}_3)\right]!}{\left[\frac{1}{2}(L^{}_3-L^{\prime}_3)\right]!} \, \begin{pmatrix} {L^{}_3}/{2} & {L^{}_3}/{2}& L^{\prime}_3 \\ n^\prime & -m^{}_{32} & m^{\prime}_3 \end{pmatrix} \nonumber\\
&& \qquad\times\; (-1)^{m^{\prime\prime}} \left[\frac{(2L_1^\prime+1)(2L_3^\prime+1)(2L^{\prime\prime}+1)}{4\pi}\right]^{1/2} \; \begin{pmatrix} L^\prime_1 & L^\prime_3 & L^{\prime\prime} \\ m_1^\prime & m_3^\prime & -m^{\prime\prime} \end{pmatrix}\, \begin{pmatrix} L^\prime_1 & L^\prime_3 & L^{\prime\prime} \\ 0 & 0 & 0 \end{pmatrix}.
\end{eqnarray}

Now, we have the expansion of each of the ${\cal{A}}$, ${\cal{B}}$, and ${\cal{C}}$ terms. To obtain the explicit form of the matrix elements, or equivalently, the detailed expansion of \eqref{matelc10}, it suffices to combine ${\cal{A}}$, ${\cal{B}}$, and ${\cal{C}}$. In this regard, we invoke Eq.~\eqref{HYHY3j}, and then we obtain:
\begin{align}\label{Matrix ele s=0}
&\sqrt{a_{\varsigma,l,k}}\, \big[\det(-\overline{b}\pz + \overline{a})\big]^{-\varsigma}\, \, {\big((g^{-1}\diamond \pz)\cdot (g^{-1}\diamond \pz)\big)^{k}\,} \, \mathcal{Y}_{l-2k,m}\big(g^{-1}\diamond \pz\big) \hspace{5cm}\nonumber\\
& \hspace{5cm} = \sum_{l^{\prime\prime\prime} k^{\prime\prime\prime} m^{\prime\prime\prime}} \textbf{U}^{(\varsigma,0)}_{l, k, m ; l^{\prime\prime\prime}, k^{\prime\prime\prime}, m^{\prime\prime\prime}} (g) \;\; \sqrt{a_{\varsigma,l^{\prime\prime\prime},k^{\prime\prime\prime}}}\, (\pz\cdot\pz)^{k^{\prime\prime\prime}}\mathcal{Y}_{l^{\prime\prime\prime}-2k^{\prime\prime\prime},m^{\prime\prime\prime}}(\pmb{z})\,,
\end{align}
where  $\mathbf{U}^{(\varsigma,0)}_{l, k, m ; l^{\prime\prime\prime}, k^{\prime\prime\prime}, m^{\prime\prime\prime}} (g) $ are the matrix elements of the Sp$(4,\R)$ scalar representations $U^{(\varsigma,0)}$ such that:
\begin{eqnarray}\label{Udefinition}
\textbf{U}^{(\varsigma,0)}_{l, k, m ; l^{\prime\prime\prime}, k^{\prime\prime\prime}, m^{\prime\prime\prime}} (g) &=& \sum_{\mathfrak{l}=0}^{\infty}\sum_{\mathfrak{k} = 0}^{\lfloor \frac{\mathfrak{l}}{2}\rfloor} \sum_{\mathfrak{m}=2\mathfrak{k}-\mathfrak{l}}^{\mathfrak{l}-2\mathfrak{k}} \sum_{\mathfrak{\underline{l}}=0}^{2k}\sum_{\mathfrak{\underline{k}} = 0}^{\lfloor \frac{\mathfrak{\underline{l}}}{2}\rfloor} \sum_{\mathfrak{\underline{m}}=2\mathfrak{\underline{k}}-\mathfrak{\underline{l}}}^{\mathfrak{\underline{l}}-2\mathfrak{\underline{k}}} \sum_{\stackrel{L^\prime L^{\prime\prime}m^{\prime\prime}}{L^\prime-L^{\prime\prime} = \mathrm{even}}} \sum_{\mathfrak{\underline{\underline{L}}}}
\frac{\sqrt{a_{\varsigma,l,k}} \,\, a_{(\varsigma,k),\mathfrak{l},\mathfrak{k}} \,\, a^\prime_{k,\mathfrak{\underline{l}},\mathfrak{\underline{k}}}}{\sqrt{a_{\varsigma,l^{\prime\prime\prime},k^{\prime\prime\prime}}}} \,\, U_{L, k, m ; L^\prime, L^{\prime\prime}, m^{\prime\prime}} (g)\nonumber\\
&& \times \; \delta^{}_{l^{\prime\prime\prime},\mathfrak{l}+\mathfrak{\underline{l}}+L^\prime} \; \delta^{}_{2k^{\prime\prime\prime},l^{\prime\prime\prime}-L^{\prime\prime\prime}} \; \big[\det\overline{a}\big]^{-\varsigma-k +1} \; \big[\det b \big]^k \; \overline{(\textbf{w}\cdot\textbf{w})}^{\mathfrak{k}}\,\, \overline{\mathcal{Y}_{\mathfrak{l}-2\mathfrak{k},\mathfrak{m}}(\textbf{w})}\,\, \overline{(\textbf{u}\cdot\textbf{u})}^{\mathfrak{\underline{k}}}\,\, \overline{\mathcal{Y}_{\mathfrak{\underline{l}}-2\mathfrak{\underline{k}},\mathfrak{\underline{m}}}(\textbf{u})} \nonumber\\
&& \times \; (-1)^{\mathfrak{\underline{\underline{m}}}} \left[\frac{(2\mathfrak{l}-4\mathfrak{k}+1)(2\mathfrak{\underline{l}}-4\mathfrak{\underline{k}}+1) (2\mathfrak{\underline{\underline{L}}}+1)}{4\pi}\right]^{1/2} \begin{pmatrix} \mathfrak{l}-2\mathfrak{k} & \mathfrak{\underline{l}}-2\mathfrak{\underline{k}} & \mathfrak{\underline{\underline{L}}} \\ \mathfrak{m} & \mathfrak{\underline{m}} & -\mathfrak{\underline{\underline{m}}} \end{pmatrix}\, \begin{pmatrix} \mathfrak{l}-2\mathfrak{k} & \mathfrak{\underline{l}}-2\mathfrak{\underline{k}} & \mathfrak{\underline{\underline{L}}} \\ 0 & 0 & 0 \end{pmatrix} \nonumber\\
&& \times \; (-1)^{m^{\prime\prime\prime}} \left[\frac{(2\mathfrak{\underline{\underline{L}}}+1)(2L^{\prime\prime}+1)(2L^{\prime\prime\prime}+1)}{4\pi}\right]^{1/2} \begin{pmatrix} \mathfrak{\underline{\underline{L}}} & L^{\prime\prime} & L^{\prime\prime\prime} \\ \mathfrak{\underline{\underline{m}}} & m^{\prime\prime} & -m^{\prime\prime\prime} \end{pmatrix}\,\begin{pmatrix} \mathfrak{\underline{\underline{L}}} & L^{\prime\prime} & L^{\prime\prime\prime} \\ 0 & 0 & 0 \end{pmatrix}.
\end{eqnarray}
Note that the definition of $a_{\varsigma,l^{\prime\prime\prime},k^{\prime\prime\prime}}$ can be understood from that of $a_{\varsigma+s,l,k}$ (already given in Eq.~\eqref{dddddd}), when we set $s=0$ and substitute $l$ and $k$ respectively by $l^{\prime\prime\prime}$ and $k^{\prime\prime\prime}$.

As already pointed out, the cases involving $b=0$, i.e., all $g=\begin{pmatrix} a & 0 \\ 0 & \overline{a} \end{pmatrix}\equiv g^{}_{(b=0)}$, need to be individually treated from the very beginning, by taking parallel steps to what given above:
\begin{eqnarray}\label{b1}
\mbox{l.h.s. Eq.~\eqref{matelc10}}\Big |_{b=0}
&&= \sqrt{a_{\varsigma,l,k}}\,\, \big[\det\overline{a}\big]^{-\varsigma}\, \,\;\; {\big((g^{-1}_{(b=0)}\diamond \pz)\cdot (g^{-1}_{(b=0)}\diamond \pz)\big)^{k}\,} \,\;\; \mathcal{Y}_{l-2k,m}\big(g^{-1}_{(b=0)}\diamond \pz\big) \nonumber\\
&&= \sqrt{a_{\varsigma,l,k}}\,\, {\big[\det\overline{a}\big]^{-\varsigma-k}}\, \big[ \det a\big]^k\, (\pz\cdot\pz)^k\, \mathcal{Y}_{l-2k,m}\big(a\pz {\overline a}^{-1}\big) \nonumber\\
&&= \sqrt{a_{\varsigma,l,k}}\,\, {\big[\det\overline{a}\big]^{-\varsigma-k}}\, \big[ \det a\big]^k\, (\pz\cdot\pz)^k \sum_{\stackrel{m_1 m_2}{m_2-m_1=m}} A_{l,k,m^{}_1,m^{}_2} \, \underbrace{D_{m_1m_2}^{l/2-k}\big[\big(z^4=0,a\pz{\overline a}^{-1}\big)\big]}_{\equiv\, D_{m_1m_2}^{l/2-k}(a\pz{\overline a}^{-1})} \nonumber\\
&&= \sqrt{a_{\varsigma,l,k}}\,\, {\big[\det\overline{a}\big]^{-\varsigma-k}}\, \big[ \det a\big]^k\, (\pz\cdot\pz)^k \sum_{\stackrel{m_1 m_2}{m_2-m_1=m}} \sum_{n_1n_2} A_{l,k,m^{}_1,m^{}_2} \, D_{m_1 n_1}^{l/2-k}(a)\, D_{n_1 n_2}^{l/2-k}(\pz)\, D_{n_2 m_2}^{l/2-k}({\overline a}^{-1}) \nonumber\\
&&= \sqrt{a_{\varsigma,l,k}}\,\, {\big[\det\overline{a}\big]^{-\varsigma-k}}\, \big[ \det a\big]^k\, (\pz\cdot\pz)^k \sum_{\stackrel{m_1 m_2}{m_2-m_1=m}} \sum_{n_1n_2} A_{l,k,m^{}_1,m^{}_2} \, D_{m_1 n_1}^{l/2-k}(a)\, D_{n_2 m_2}^{l/2-k}({\overline a}^{-1}) \nonumber\\
&& \qquad\times\; \sum_{\substack{l^{\prime}\\(l-2k)-l^{\prime} = \mathrm{even}\\ (\mathrm{then,}\,l-l^\prime=\mathrm{even}\equiv 2k^\prime)}} (4\pi)^{1/2}\, (-1)^{l^{\prime}+n_2}\; 2^{l^{\prime}}\; (2l^{\prime}+1)^{1/2} \left(\frac{\big((l-2k)-l^{\prime}\big)!}{\big((l-2k)+l^{\prime}+1\big)!}\right)^{1/2} \nonumber\\
&&\hspace{3.25cm}\times\; \frac{\left[\frac{1}{2}\big((l-2k)+l^{\prime}\big)\right]!}{\left[\frac{1}{2}\big((l-2k)-l^{\prime}\big)\right]!}\,  \begin{pmatrix} \frac{(l-2k)}{2} & \frac{(l-2k)}{2}& l^{\prime} \\ n_1 & -n_2 & m^{\prime} \end{pmatrix} (\pz\cdot\pz)^{\frac{(l-2k)-l^{\prime}}{2}}\mathcal{Y}_{l^{\prime}m^{\prime}}(\pmb{z})\,.
\end{eqnarray}
By adjusting the above expansion, we find the corresponding matrix elements such that:
\begin{eqnarray}\label{b2}
\mbox{l.h.s. Eq.~\eqref{matelc10}}\Big |_{b=0} 
&=& \sum_{k^\prime m^\prime} \frac{\sqrt{a_{\varsigma,l,k}}}{\sqrt{a_{\varsigma,l,k^\prime}}}\, {\big[\det\overline{a}\big]^{-\varsigma-k}}\, \big[ \det a\big]^k\, \sum_{\stackrel{m_1 m_2}{m_2-m_1=m}} \sum_{n_1n_2} A_{l,k,m^{}_1,m^{}_2} \, D_{m_1 n_1}^{l/2-k}(a)\, D_{n_2 m_2}^{l/2-k}({\overline a}^{-1}) \nonumber\\
&&\times\; (4\pi)^{1/2}\, (-1)^{l-2k^{\prime}+n_2}\; 2^{l-2k^{\prime}}\; \big(2(l-2k^{\prime})+1\big)^{1/2} \left(\frac{\big(2k^{\prime}-2k\big)!}{\big(2l-2k-2k^{\prime}+1\big)!}\right)^{1/2} \nonumber\\
&&\times\; \frac{\left[\frac{1}{2}\big(2l-2k-2k^{\prime}\big)\right]!}{\left[\frac{1}{2}\big(2k^{\prime}-2k\big)\right]!}\,  \begin{pmatrix} \frac{(l-2k)}{2} & \frac{(l-2k)}{2}& l-2k^{\prime} \\ n_1 & -n_2 & m^{\prime} \end{pmatrix} \sqrt{a_{\varsigma,l,k^\prime}} \, (\pz\cdot\pz)^{k^{\prime}}\mathcal{Y}_{l-2k^{\prime},m^{\prime}}(\pmb{z}) \nonumber\\
&\equiv& \sum_{k^\prime m^\prime} \textbf{U}^{(\varsigma,0)}_{l, k, m ; l, k^{\prime}, m^{\prime}} \big(g^{}_{(b=0)}\big) \;\; \sqrt{a_{\varsigma,l,k^\prime}}\, (\pz\cdot\pz)^{k^{\prime}}\mathcal{Y}_{l-2k^{\prime},m^{\prime}}(\pmb{z})\,.
\end{eqnarray}
Hence:
\begin{eqnarray}\label{b3}
 \textbf{U}^{(\varsigma,0)}_{l, k, m ; l, k^{\prime}, m^{\prime}} \big(g^{}_{(b=0)}\big) 
&=&  \frac{\sqrt{a_{\varsigma,l,k}}}{\sqrt{a_{\varsigma,l,k^\prime}}}\, {\big[\det\overline{a}\big]^{-\varsigma-k}}\, \big[ \det a\big]^k\, \sum_{\stackrel{m_1 m_2}{m_2-m_1=m}} \sum_{n_1n_2} A_{l,k,m^{}_1,m^{}_2} \, D_{m_1 n_1}^{l/2-k}(a)\, D_{n_2 m_2}^{l/2-k}({\overline a}^{-1}) \nonumber\\
&&\times\; (4\pi)^{1/2}\, (-1)^{l-2k^{\prime}+n_2}\; 2^{l-2k^{\prime}}\; \big(2(l-2k^{\prime})+1\big)^{1/2} \left(\frac{\big(2k^{\prime}-2k\big)!}{\big(2l-2k-2k^{\prime}+1\big)!}\right)^{1/2} \nonumber\\
&&\times\; \frac{\left[\frac{1}{2}\big(2l-2k-2k^{\prime}\big)\right]!}{\left[\frac{1}{2}\big(2k^{\prime}-2k\big)\right]!}\,  \begin{pmatrix} \frac{(l-2k)}{2} & \frac{(l-2k)}{2}& l-2k^{\prime} \\ n_1 & -n_2 & m^{\prime} \end{pmatrix} \,. 
\end{eqnarray}

\subsection{The general case $U^{(\varsigma,s)}$}
The general case corresponds to that shown in Eq. \eqref{UIRAdS}. The orthonormal basis of the (Segal-)Bargmann-Fock Hilbert space $\mathcal{F}^{(\vsi,s)}$ can be given in terms of the set of holomorphic functions:
\begin{eqnarray}
\sqrt{a_{\varsigma+s,l,k}}\, (\pz\cdot\pz)^{k}\, \mathcal{Y}_{s,l-2k,JM}(\pz) = \sum_{m \rho} (s\rho ,l-2k, m| s ,l-2k, J M)\,e_{s\rho}\, \sqrt{a_{\varsigma+s,l,k}}\, (\pz\cdot\pz)^{k}\,\mathcal{Y}_{l-2k,m}(\pz)\,,
\end{eqnarray}
where the allowed ranges of the parameters and the definition of $a_{\varsigma+s,l,k}$ can be found, respectively, in Eqs. \eqref{nkm1m2} and \eqref{dddddd}. On the other hand, the Clebsch-Gordan coefficients are defined by \eqref{3japc}, and $e_{s\rho}$, with $-s\leqslant \rho \leqslant s$, by \eqref{essig}.

Proceeding as before, considering this basis along with the action \eqref{UIRAdS}, the matrix elements of the (unitary or not) representation $U^{(\varsigma,s)}$ are obtained by expanding the following formula:
\begin{align}\label{spinUIR}
&\sqrt{a_{\varsigma+s,l,k}}\, \big[\det(-\overline{b}\pz + \overline{a})\big]^{-\varsigma-s}\, D^s(\pz b^{\ast} + a^{\ast}) \,{\big((g^{-1}\diamond \pz)\cdot (g^{-1}\diamond \pz)\big)^{k}\,} \,\; \mathcal{Y}_{s,l-2k,JM}\big(g^{-1}\diamond \pz\big) \nonumber\\
&\hspace{2cm} = \sum_{m\rho} (s\rho ,l-2k, m| s ,l-2k, J M)\, \underbrace{D^s(\pz b^{\ast} + a^{\ast})}_{\equiv{\cal{A}}^\prime} \nonumber\\
&\hspace{3cm} \times \,e_{s\rho}\, \underbrace{\sqrt{a_{\varsigma+s,l,k}}\, \big[\det(-\overline{b}\pz + \overline{a})\big]^{-\varsigma-s}\, \,\; {\big((g^{-1}\diamond \pz)\cdot (g^{-1}\diamond \pz)\big)^{k}\,} \,\; \mathcal{Y}_{l-2k,m}\big(g^{-1}\diamond \pz\big)}_{\equiv{\cal{B}}^\prime}\,,
\end{align}
which in turn requires the expansion of the terms ${\cal{A}}^\prime$ and ${\cal{B}}^\prime$.
\medskip

\emph{\textbf{Term ${\cal{A}}^\prime$.}} We again invoke the relations \eqref{addexp1} and \eqref{multexp}, based upon which the following expansion formula holds:
\begin{align}\label{Aprime}
{\cal{A}}^\prime &\equiv D^s(\pz b^{\ast} + a^{\ast}) = \sum_{n_1 n_2} e^{}_{sn_1}\, D^s_{n_1 n_2}(\pz b^{\ast} + a^{\ast})\, e^{\dagger}_{sn_2}\nonumber\\
&= \sum_{n_1 n_2} e^{}_{sn_1}\,e^{\dagger}_{sn_2}\, \left( \sigma_{n_1 ,n_2}^s \right)^{-1} \sum_{s^\prime n^\prime_1 n^\prime_2} \sum_{n^\prime} \sigma_{n_1-n^\prime_1\, n_2 -n^\prime_2}^{s-s^\prime} \, D^{s-s^\prime}_{n^\prime,n_2 -n^\prime_2} (b^{\ast}) \,\sigma_{n^\prime_1, n^\prime_2}^{s'} \,D_{n^\prime_1 n^\prime_2}^{s^\prime} (a^{\ast}) \,D^{s-s^\prime}_{n_1-n_1^\prime, n^\prime} (\pz)\,,
\end{align}
with $-s\leqslant n_1, n_2 \leqslant s$. On the other hand, having Eq.~\eqref{DlYl2} in mind, we obtain:
\begin{eqnarray}\label{Aprime'}
{\cal{A}}^\prime \equiv D^s(\pz b^{\ast} + a^{\ast}) &=& \sum_{n_1 n_2} e^{}_{sn_1}\,e^{\dagger}_{sn_2}\, \left( \sigma_{n_1 ,n_2}^s \right)^{-1} \sum_{s^\prime n^\prime_1 n^\prime_2} \sum_{n^\prime} \sum_{\stackrel{s^{\prime\prime}}{2(s-s^{\prime})-s^{\prime\prime} = \mathrm{even}}} \sigma_{n_1-n^\prime_1\, n_2 -n^\prime_2}^{s-s^\prime} \, D^{s-s^\prime}_{n^\prime,n_2 -n^\prime_2} (b^{\ast}) \,\sigma_{n^\prime_1, n^\prime_2}^{s'} \,D_{n^\prime_1 n^\prime_2}^{s^\prime} (a^{\ast}) \nonumber\\
&& \quad\quad \times \, (4\pi)^{1/2}\, (-1)^{s^{\prime\prime}+n^\prime}\; 2^{s^{\prime\prime}}\; (2s^{\prime\prime}+1)^{1/2} \left(\frac{\big(2(s-s^{\prime})-s^{\prime\prime}\big)!}{\big(2(s-s^{\prime})+s^{\prime\prime}+1\big)!}\right)^{1/2} \frac{\left[\frac{1}{2}\big(2(s-s^{\prime})+s^{\prime\prime}\big)\right]!}{\left[\frac{1}{2}\big(2(s-s^{\prime})-s^{\prime\prime}\big)\right]!} \nonumber\\
&& \quad\quad \times \, \begin{pmatrix} s-s^\prime & s-s^\prime & s^{\prime\prime} \\ n_1-n_1^\prime & -n^\prime & n^{\prime\prime} \end{pmatrix} (\pz\cdot\pz)^{\frac{2(s-s^{\prime})-s^{\prime\prime}}{2}}\mathcal{Y}_{s^{\prime\prime}n^{\prime\prime}}(\pmb{z})\,.
\end{eqnarray}

\emph{\textbf{Term ${\cal{B}}^\prime$.}} With reference to Eq.~\eqref{Matrix ele s=0}, we have:
\begin{align}\label{Bprime}
&{\cal{B}}^\prime \equiv \sqrt{a_{\varsigma+s,l,k}}\, \big[\det(-\overline{b}\pz + \overline{a})\big]^{-\varsigma-s}\, \, {\big((g^{-1}\diamond \pz)\cdot (g^{-1}\diamond \pz)\big)^{k}\,} \, \mathcal{Y}_{l-2k,m}\big(g^{-1}\diamond \pz\big) \hspace{5cm}\nonumber\\
& \hspace{5cm} = \sum_{l^{\prime\prime\prime} k^{\prime\prime\prime} m^{\prime\prime\prime}} \textbf{U}^{(\varsigma+s,0)}_{l, k, m ; l^{\prime\prime\prime}, k^{\prime\prime\prime}, m^{\prime\prime\prime}} (g) \; \sqrt{a_{\varsigma+s,l^{\prime\prime\prime},k^{\prime\prime\prime}}}\, (\pz\cdot\pz)^{k^{\prime\prime\prime}}\mathcal{Y}_{l^{\prime\prime\prime}-2k^{\prime\prime\prime},m^{\prime\prime\prime}}(\pmb{z})\,,
\end{align}
where the definition of the new entity $\textbf{U}^{(\varsigma+s,0)}_{l, k, m ; l^{\prime\prime\prime}, k^{\prime\prime\prime}, m^{\prime\prime\prime}} (g)$ can be understood from that of $\textbf{U}^{(\varsigma,0)}_{l, k, m ; l^{\prime\prime\prime}, k^{\prime\prime\prime}, m^{\prime\prime\prime}} (g)$ (i.e., \eqref{Udefinition}), when $\varsigma$ is replaced by $\varsigma+s$.

Now, we need to combine the above results to have the explicit expansion of \eqref{spinUIR}. To do so, we first use Eq. \eqref{HYHY3j} and then \eqref{converse}. Accordingly, we have:
\begin{align}\label{Matrix ele}
&\sqrt{a_{\varsigma+s,l,k}}\, \big[\det(-\overline{b}\pz + \overline{a})\big]^{-\varsigma-s}\, D^s(\pz b^{\ast} + a^{\ast}) \,{\big((g^{-1}\diamond \pz)\cdot (g^{-1}\diamond \pz)\big)^{k}\,} \,\; \mathcal{Y}_{s,l-2k,JM}\big(g^{-1}\diamond \pz\big) \nonumber\\
&\hspace{2cm} = \sum_{\underline{l}^{\prime\prime\prime}\, \underline{k}^{\prime\prime\prime}\, \underline{J}^{\prime\prime\prime}\, \underline{M}^{\prime\prime\prime}} \textbf{U}^{(\varsigma,s)}_{l,k,J,M; \underline{l}^{\prime\prime\prime},\underline{k}^{\prime\prime\prime}, \underline{J}^{\prime\prime\prime},\underline{M}^{\prime\prime\prime}} (g) \, \sqrt{a_{\varsigma+s,\underline{l}^{\prime\prime\prime},\underline{k}^{\prime\prime\prime}}}\, (\pz\cdot\pz)^{\underline{k}^{\prime\prime\prime}}\; \mathcal{Y}_{s,\underline{l}^{\prime\prime\prime}-2\underline{k}^{\prime\prime\prime}, \underline{J}^{\prime\prime\prime}\, \underline{M}^{\prime\prime\prime}}(\pz)\,,
\end{align}
with:
\begin{align}\label{smatrixelements}
&\textbf{U}^{(\varsigma,s)}_{l,k,J,M; \underline{l}^{\prime\prime\prime},\underline{k}^{\prime\prime\prime}, \underline{J}^{\prime\prime\prime},\underline{M}^{\prime\prime\prime}} (g) = \sum_{n_1 n_2} e^{}_{sn_1}\,e^{\dagger}_{sn_2}\, \left( \sigma_{n_1 ,n_2}^s \right)^{-1} \sum_{m\,\rho} \sum_{s^\prime n^\prime_1 n^\prime_2} \sum_{n^\prime} \sum_{\stackrel{s^{\prime\prime}}{2(s-s^{\prime})-s^{\prime\prime} = \mathrm{even}}} \sum_{l^{\prime\prime\prime} k^{\prime\prime\prime} m^{\prime\prime\prime}} \nonumber\\
& \quad\quad \times \, \frac{\sqrt{a_{\varsigma+s,l^{\prime\prime\prime},k^{\prime\prime\prime}}}} {\sqrt{a_{\varsigma+s,\underline{l}^{\prime\prime\prime},\underline{k}^{\prime\prime\prime}}}}\, \delta_{\underline{l}^{\prime\prime\prime},{l}^{\prime\prime\prime}+2(s-s^\prime)}\, \delta_{2\underline{k}^{\prime\prime\prime},\underline{l}^{\prime\prime\prime} - \underline{L}^{\prime\prime\prime}}\, \sigma_{n_1-n^\prime_1\, n_2 -n^\prime_2}^{s-s^\prime} \, D^{s-s^\prime}_{n^\prime,n_2 -n^\prime_2} (b^{\ast}) \,\sigma_{n^\prime_1 ,n^\prime_2}^{s'} \,D_{n^\prime_1 n^\prime_2}^{s^\prime} (a^{\ast}) \nonumber\\
& \quad\quad \times \, (4\pi)^{1/2}\, (-1)^{s^{\prime\prime}+n^\prime}\; 2^{s^{\prime\prime}}\; (2s^{\prime\prime}+1)^{1/2} \, \left(\frac{\big(2(s-s^{\prime})-s^{\prime\prime}\big)!}{\big(2(s-s^{\prime})+s^{\prime\prime}+1\big)!}\right)^{1/2} \frac{\left[\frac{1}{2}\big(2(s-s^{\prime})+s^{\prime\prime}\big)\right]!}{\left[\frac{1}{2}\big(2(s-s^{\prime})-s^{\prime\prime}\big)\right]!} \, \begin{pmatrix} s-s^\prime & s-s^\prime & s^{\prime\prime} \\ n_1-n_1^\prime & -n^\prime & n^{\prime\prime} \end{pmatrix} \nonumber\\
& \quad\quad \times \, \textbf{U}^{(\varsigma+s,0)}_{l, k, m ; l^{\prime\prime\prime}, k^{\prime\prime\prime}, m^{\prime\prime\prime}} (g)\, (-1)^{\underline{m}^{\prime\prime\prime}} \left[\frac{(2s^{\prime\prime}+1)(2l^{\prime\prime\prime}-4k^{\prime\prime\prime}+1)(2\underline{L}^{\prime\prime\prime}+1)}{4\pi}\right]^{1/2} \, \begin{pmatrix} s^{\prime\prime} & l^{\prime\prime\prime}-2k^{\prime\prime\prime} & \underline{L}^{\prime\prime\prime} \\ n^{\prime\prime} & m^{\prime\prime\prime} & -\underline{m}^{\prime\prime\prime} \end{pmatrix}\nonumber\\
& \quad\quad \times \,\begin{pmatrix} s^{\prime\prime} & l^{\prime\prime\prime}-2k^{\prime\prime\prime} & \underline{L}^{\prime\prime\prime} \\ 0 & 0 & 0 \end{pmatrix}\, (s\rho ,l-2k, m| s ,l-2k, J M)\, \overline{(s\rho^\prime ,\underline{l}^{\prime\prime\prime}-2\underline{k}^{\prime\prime\prime}, \underline{m}^{\prime\prime\prime}| s ,\underline{l}^{\prime\prime\prime}-2\underline{k}^{\prime\prime\prime}, \underline{J}^{\prime\prime\prime}\, \underline{M}^{\prime\prime\prime})}\, e^{}_{s\rho}\, e^{\dagger}_{s\rho^\prime}\,.
\end{align}

As before, it should be noted that, the cases with $g=\rm{diag}(a, \overline{a} )\equiv g^{}_{(b=0)}$ need to be individually treated from the very beginning:
\begin{eqnarray}
\mbox{l.h.s. Eq.~\eqref{spinUIR}}\Big |_{b=0} 
&=& \sqrt{a_{\varsigma+s,l,k}}\, \big[\det\overline{a}\big]^{-\varsigma-s}\, D^s(a^{\ast}) \,{\big((g^{-1}_{(b=0)}\diamond \pz)\cdot (g^{-1}_{(b=0)}\diamond \pz)\big)^{k}\,} \,\; \mathcal{Y}_{s,l-2k,JM}\big(g^{-1}_{(b=0)}\diamond \pz\big) \nonumber\\
&=& \sum_{m\rho} (s\rho ,l-2k, m| s ,l-2k, J M)\, D^s(a^{\ast}) \nonumber\\
&& \quad\times \, e_{s\rho}\, \underbrace{\sqrt{a_{\varsigma+s,l,k}}\, \big[\det\overline{a}\big]^{-\varsigma-s}\, {\big((g^{-1}_{(b=0)}\diamond \pz)\cdot (g^{-1}_{(b=0)}\diamond \pz)\big)^{k}\,} \, \mathcal{Y}_{l-2k,m}\big(g^{-1}_{(b=0)}\diamond \pz\big)}_{\mbox{see Eqs. \eqref{b1} and \eqref{b2}; $\varsigma\mapsto\varsigma+s$}} \nonumber\\
&=& \sum_{m\rho} (s\rho ,l-2k, m| s ,l-2k, J M)\, D^s(a^{\ast}) \nonumber\\
&&\quad \times \, e_{s\rho}\, \sum_{k^\prime m^\prime} \textbf{U}^{(\varsigma+s,0)}_{l, k, m ; l, k^{\prime}, m^{\prime}} \big(g^{}_{(b=0)}\big) \; \sqrt{a_{\varsigma+s,l,k^\prime}}\, (\pz\cdot\pz)^{k^{\prime}}\, \mathcal{Y}_{l-2k^{\prime},m^{\prime}}(\pmb{z})\,.
\end{eqnarray}
Employing Eq.~\eqref{converse} and also making some adjustments, we obtain the corresponding matrix elements as:
\begin{eqnarray}\label{kabab}
\mbox{l.h.s. Eq.~\eqref{spinUIR}}\Big |_{b=0} 
&=& \sum_{k^\prime J^\prime M^\prime} \sum_{m\rho} \sum_{m^\prime} \textbf{U}^{(\varsigma+s,0)}_{l, k, m ; l, k^{\prime}, m^{\prime}} \big(g^{}_{(b=0)}\big)\, D^s(a^{\ast}) \nonumber\\
&&\quad \times\; (s\rho ,l-2k, m| s ,l-2k, J M)\, \overline{(s\rho^\prime, l-2k^\prime, m^\prime| s, l-2k^\prime, J^\prime M^\prime)}\, e_{s\rho}\, e_{s\rho^\prime}^\dagger \nonumber\\
&&\quad \times\; \sqrt{a_{\varsigma+s,l,k^\prime}}\, (\pz\cdot\pz)^{k^{\prime}}\, \mathcal{Y}_{s,l-2k^\prime,J^\prime M^\prime}(\pz)\nonumber\\
&\equiv& \sum_{k^\prime J^\prime M^\prime} \textbf{U}^{(\varsigma,s)}_{l, k, J, M ; l, k^{\prime}, J^\prime, M^{\prime}} \big(g^{}_{(b=0)}\big)\, \sqrt{a_{\varsigma+s,l,k^\prime}}\, (\pz\cdot\pz)^{k^{\prime}}\, \mathcal{Y}_{s,l-2k^\prime,J^\prime M^\prime}(\pz)\,,
\end{eqnarray}
whose explicit expression is:
\begin{eqnarray}\label{kabab1}
 \textbf{U}^{(\varsigma,s)}_{l, k, J, M ; l, k^{\prime}, J^\prime, M^{\prime}} \big(g^{}_{(b=0)}\big)
&=& \sum_{m\rho} \sum_{m^\prime} \textbf{U}^{(\varsigma+s,0)}_{l, k, m ; l, k^{\prime}, m^{\prime}} \big(g^{}_{(b=0)}\big)\, D^s(a^{\ast}) \nonumber\\
&&\quad \times\; (s\rho ,l-2k, m| s ,l-2k, J M)\, \overline{(s\rho^\prime, l-2k^\prime, m^\prime| s, l-2k^\prime, J^\prime M^\prime)}\, e_{s\rho}\, e_{s\rho^\prime}^\dagger
\end{eqnarray}

\setcounter{equation}{0} 
\section{Computation of characters} \label{Sec. characters}

\subsection{Eigenvalues of $g\in\mathrm{Sp}(4,\R)$}
Finding the eigenvalues of $g\in\mathrm{Sp}(4,\R)$ (see Eq. \eqref{sp4Rdef}) amounts to solve $\det\begin{pmatrix} a-\lambda & b \\ -\overline b & \overline a - \lambda \end{pmatrix} = 0$, for which, according to the identities given in Eq. \eqref{det}, we have four equivalent equations:
\begin{eqnarray}
\label{1} \big(\det(a-\lambda)\big) \det ((\overline a - \lambda) + \overline{b} (a - \lambda)^{-1}b) &=& 0\,,\\[0.2cm]
\label{2} \big(\det(\overline a-\lambda)\big) \det ((a-\lambda) +b(\overline a - \lambda)^{-1}{\overline{b}}) &=& 0\,,\\[0.2cm]
\label{3} \big(\det b\big) \det (\overline b +(\overline a - \lambda)b^{-1}(a-\lambda)) &=& 0\,,\\[0.2cm]
\label{4} \big(\det \overline b\big) \det (b +(a-\lambda){\overline{b}}^{-1}(\overline a - \lambda)) &=& 0\,.
\end{eqnarray}
From the above, it is evident that if $\lambda=\mu\; \big(\in\mathbb{C}\big)$ is a root, than $\lambda=\overline{\mu}$ is a root as well.\footnote{Otherwise, Eq. \eqref{1} is not equivalent with its complex conjugate twin \eqref{2}, and similarly, Eq. \eqref{3} with \eqref{4}. [Note that this reasoning is also consistent with the diagonal elements (namely, $a$ and its complex conjugate $\overline{a}$) of the matrix $g\in\mathrm{Sp}(4,\R)$ (see Eq. \eqref{sp4Rdef}).]} A complete realization of the eigenvalues, however, can be obtained by considering the diagonalized form of the generic member $g$ of the group, in terms of its eigenvalues, when extended to SL$(4,\mathbb{C})$. To achieve this, while keeping Eq. \eqref{sp4Rdef} and its subsequent identities in mind, we use the notations introduced in Appendix \ref{App. quaternions} (specifically, Eq. \eqref{cqumat}). Thus, we have the diagonalized form of $g$:
\begin{align}\label{gd}
 g^{}_d &= \begin{pmatrix} a^{}_d\equiv (z^4,z^3\pmb{e}_3) & 0 \\ 0 & {\overline{a}}^{}_d\equiv\overline{({z^4},{z^3}\pmb{e}_3)} \end{pmatrix} \nonumber\\[0.2cm]
&=
\begin{pmatrix}
  \begin{pmatrix}
   z^4+\ii z^3 \equiv\mu & 0 \\
    0 & z^4-\ii z^3\equiv\nu
   \end{pmatrix} & 0 \\
   0 & \begin{pmatrix}
   \overline{z^4}+\ii \overline{z^3}\equiv\overline{\nu} & 0 \\
    0 & \overline{z^4}-\ii \overline{z^3}\equiv\overline{\mu}
\end{pmatrix}
\end{pmatrix},
\end{align}
where, by construction, $\mu,\nu\in\mathbb{C}$ and $|\mu|=1/|\nu|$. The latter identity can be easily understood, once we recall that $\det(g^{}_d) = 1$. In summary, the above arguments reveal that the four eigenvalues of $g\in\mathrm{Sp}(4,\R)$ are of the forms $\lambda=\mu,\nu,\overline{\nu}$, and $\overline{\mu}$, with the above definitions and constraints.

\subsection{Characters of the scalar representations $U^{(\vsi,s=0)}$}

For a generic element $g\in\mathrm{Sp}(4,\R)$ of the form $\begin{pmatrix} a=(z^4,\pz) & 0 \\ 0 & \overline{a}=\overline{({z^4},{\pz})} \end{pmatrix}\equiv g^{}_{(b=0)}$, such that $\pz$ is not necessarily $z^3\pmb{e}_3$, the matrix elements of the corresponding scalar representations $U^{(\vsi,0)}$ have been already given in Eq. \eqref{b2}. Let us rewrite them for convenience:
\begin{eqnarray}\label{Scalar matrix ele 01}
\textbf{U}^{(\varsigma,0)}_{l, k, m ; l, k^\prime, m^\prime} (g^{}_{(b=0)}) &=& \frac{\sqrt{a_{\varsigma,l,k}}}{\sqrt{a_{\varsigma,l,k^\prime}}}\, {\big[\det\overline{a}\big]^{-\varsigma-k}}\, \big[\det a\big]^k\, \sum_{\stackrel{m_1 m_2}{m_2-m_1=m}} \sum_{n_1n_2} A_{l,k,m^{}_1,m^{}_2} \, D_{m_1 n_1}^{l/2-k}(a)\, D_{n_2 m_2}^{l/2-k}({\overline a}^{-1}) \nonumber\\
&&\times\; (4\pi)^{1/2}\, (-1)^{l-2k^{\prime}+n_2}\; 2^{l-2k^{\prime}}\; \big(2(l-2k^{\prime})+1\big)^{1/2} \left(\frac{\big(2k^{\prime}-2k\big)!}{\big(2l-2k-2k^{\prime}+1\big)!}\right)^{1/2} \nonumber\\
&&\times\; \frac{\left[\frac{1}{2}\big(2l-2k-2k^{\prime}\big)\right]!}{\left[\frac{1}{2}\big(2k^{\prime}-2k\big)\right]!}\,  \begin{pmatrix} \frac{(l-2k)}{2} & \frac{(l-2k)}{2}& l-2k^{\prime} \\ n_1 & -n_2 & m^{\prime} \end{pmatrix},
\end{eqnarray}
with:
\begin{align}
A_{l,k,m_1,m_2} = 2^{-l+2k}\, \sigma^{l-2k}_{m=m_2-m_1} \, \left(\frac{2l-4k+1}{4\pi}\right)^{1/2} (-1)^{m_1} \; \big(\sigma^{l/2-k}_{m_1 ,m_2}\big)^{-1} \,.
\end{align}

In the above relation, if we further restrict $a=(z^4,\pz)$ to $a^{}_d=(z^4,z^3)= \begin{pmatrix} z^4+\ii z^3 \equiv\mu & 0 \\ 0 & z^4-\ii z^3\equiv\nu \end{pmatrix}$ (and correspondingly, $g^{}_{(b=0)}$ to $g^{}_d$), the matrix elements of the SU$(2)$ representations $D_{m_1 n_1}^{l/2-k}(a^{}_d)$ and $D_{n_2 m_2}^{l/2-k}({\overline a}^{-1}_d)$ are non-vanishing if and only if $m_1=n_1$ and $m_2=n_2$, respectively; this point becomes clearer when one considers the matrix representation of the quaternion $a^{}_d$ and closely examines Eq. \eqref{matelHC}. Adjusting $m_1=n_1$ and $m_2=n_2$, the involved $3-j$ symbol necessitates $m=m^\prime$, and subsequently, $k=k^\prime$; the latter identity guarantees that $m$ and $m^\prime$ run on the same values $2k-l\leqslant m,m^\prime \leqslant l-2k$, which is a requirement of the former identity. In summary, in the case $a=a^{}_d$, the non-vanishing matrix elements of the scalar representations $U^{(\vsi,0)}$ associated with $g^{}_d$ are only the diagonal ones and explicitly read as:
\begin{eqnarray}\label{0matrixelementgd}
\textbf{U}^{(\varsigma,0)}_{l, k, m ; l, k, m} (g^{}_d) &=& {(\overline{\nu}\, \overline{\mu})^{-\varsigma}} \sum_{\stackrel{m_1 m_2}{m_2-m_1=m}} A_{l,k,m^{}_1,m^{}_2} \, \mu^{l/2-m_1}\, \nu^{l/2+m_1} \, {\overline{\mu}}^{-l/2+m_2}\, {\overline{\nu}}^{-l/2-m_2} \nonumber\\
&&\times\; (4\pi)^{1/2}\, (-1)^{l-2k+m_2}\; 2^{l-2k}\; \left(\frac{1}{\big[2(l-2k)\big]!}\right)^{1/2} \; (l-2k)!\, \begin{pmatrix} \frac{(l-2k)}{2} & \frac{(l-2k)}{2} & l-2k \\ m_1 & -m_2 & m \end{pmatrix}.
\end{eqnarray}
Then, the characters of the representations $U^{(\vsi,0)}$ are given by: 
\begin{equation}\label{character0}
\mbox{Tr}\,\textbf{U}^{(\varsigma,0)}(g^{}_d) = \sum_{lkm} \textbf{U}^{(\varsigma,0)}_{l, k, m ; l, k, m}(g^{}_d)\,.
\end{equation}


\subsection{Characters of the spin representations $U^{(\vsi,s)}$}
According to our calculations given in the previous section (specifically, Eq. \eqref{kabab1}), the matrix elements of the spin representations $U^{(\vsi,s)}$ associated with a generic element $g\in\mathrm{Sp}(4,\R)$ of the form $\begin{pmatrix} a=(z^4,\pz) & 0 \\ 0 & \overline{a}=\overline{({z^4},{\pz})} \end{pmatrix}\equiv g^{}_{(b=0)}$, such that $\pz$ is not necessarily $z^3\pmb{e}_3$, read:
\begin{eqnarray}
\textbf{U}^{(\varsigma,s)}_{l, k, J, M ; l, k^{\prime}, J^\prime, M^{\prime}} \big(g^{}_{(b=0)}\big) &=& \sum_{m\rho} \sum_{m^\prime} \textbf{U}^{(\varsigma+s,0)}_{l, k, m ; l, k^{\prime}, m^{\prime}} \big(g^{}_{(b=0)}\big)\, \overbrace{\sum_{n_1 n_2} e_{sn_1}^{} D^s_{n_1n_2}(a^{\ast})\, e_{sn_2}^\dagger}^{D^s(a^{\ast})} \nonumber\\
&& \times\; (s\rho ,l-2k, m| s ,l-2k, J M)\, \overline{(s\rho^\prime, l-2k^\prime, m^\prime| s, l-2k^\prime, J^\prime M^\prime)}\, e_{s\rho}\, e_{s\rho^\prime}^\dagger \,,
\end{eqnarray}
where the definition of $\textbf{U}^{(\varsigma+s,0)}_{l, k, m ; l, k^{\prime}, m^{\prime}} \big(g^{}_{(b=0)}\big)$ can be understood from that of $\textbf{U}^{(\varsigma,0)}_{l, k, m ; l, k^\prime, m^\prime} (g^{}_{(b=0)})$ (see Eq. \eqref{Scalar matrix ele 01}), when $\varsigma\mapsto\varsigma+s$. Proceeding as above, if we now further restrict $a=(z^4,\pz)$ to $a^{}_d=(z^4,z^3)= \begin{pmatrix} z^4+\ii z^3 \equiv\mu & 0 \\ 0 & z^4-\ii z^3\equiv\nu \end{pmatrix}$ (and correspondingly, $g^{}_{(b=0)}$ to $g^{}_d$), the matrix elements become:
\begin{eqnarray}\label{smatrixelementgd}
\textbf{U}^{(\varsigma,s)}_{l, k, J, M ; l, k, J^\prime, M^{\prime}} \big(g^{}_{d}\big) &=& \sum_{m\rho} \textbf{U}^{(\varsigma+s,0)}_{l, k, m ; l, k, m} \big(g^{}_{d}\big)\, \sum_{n\;(=n_1=n_2)} (\overline{\mu})^{s-n}\, (\overline{\nu})^{s+n} \, e_{sn}^{}\, e_{sn}^\dagger \nonumber\\
&& \times\; (s\rho ,l-2k, m| s ,l-2k, J M)\, \overline{(s\rho^\prime, l-2k, m| s, l-2k, J^\prime M^\prime)}\, e_{s\rho}^{}\, e_{s\rho^\prime}^\dagger\,.
\end{eqnarray}
Subsequently, the characters of the representations $U^{(\vsi,s)}$ are: 
\begin{equation}\label{characters}
\mbox{Tr}\,\textbf{U}^{(\varsigma,s)}(g^{}_d) = \sum_{l, k, J, M} \textbf{U}^{(\varsigma,s)}_{l, k, J, M ; l, k, J, M}(g^{}_d)\,.
\end{equation}

\setcounter{equation}{0} \section{Summary} \label{Sec.:Summary}

This paper completes  Ref.~\cite{AdS00},  where we studied among other issues the ``massive''  UIRs of Sp$(4,\R)$, i.e., those UIRs that contract to the Poincar\'{e} massive ($m^2 >0$) UIRs. Here we present mathematical aspects of these UIRs relatives to the elements of matrix of these infinite dimensional representations as well as their characters.  These results are fundamental to be able to apply covariant quantization \cite{gazeau2014,gazeau2013,gazeau2016,gazeau2016a} to the massive anti-de Siiter classical systems following the framework developed in Ref. \cite{olmogaz19} for the  $(1+1)$  anti-de Sitter group. The results will be published elsewhere.
\medskip

The main results of the paper are the following ones:
\begin{enumerate}

\item 
For the scalar UIRs of Sp$(4,\R)$, $U^{(\varsigma,0)}$,  the matrix elements, $\mathbf{U}^{(\varsigma,0)}_{l, k, m ; l^{\prime\prime\prime}, k^{\prime\prime\prime}, m^{\prime\prime\prime}} (g) $, are given by \eqref{Udefinition} in the orthogonal basis of of the (Segal-)Bargmann-Fock Hilbert space $\mathcal{F}^{(\vsi,0)}$
determined by the functions 
 $\mathrm{F}^{(\varsigma,0)}_{l,k,m}(\pz)$ \eqref{havich}.

In the particular case that the elements $g^{}_{(b=0)}$ of Sp$(4,\R)$, i.e. those  have  $b=0$ (see \eqref{sp4Rdef}) we find that the matrix elements, 
$\textbf{U}^{(\varsigma,0)}_{l, k, m ; l, k^{\prime}, m^{\prime}} \big(g^{}_{(b=0)}\big)$  are expressed  by \eqref{b3}.
 
The character of the representations $\mathbf{U}^{(\varsigma,0)}$ are given by $\mbox{Tr}\,\textbf{U}^{(\varsigma,0)}(g^{}_d) = \sum_{lkm} \textbf{U}^{(\varsigma,0)}_{l, k, m ; l, k, m}(g^{}_d)$ \eqref{character0},  where $\textbf{U}^{(\varsigma,0)}_{l, k, m ; l, k, m}(g^{}_d)$ is expressed by \eqref{0matrixelementgd} and 
$g^{}_d$ is the diagonalized form of the generic element 
$g\in $ Sp$(4,\R)$  in terms of its eigenvalues, when extended to SL$(4,\mathbb{C})$ \eqref{gd}.

\item
For the general case $U^{(\varsigma,s)}$,   i.e. spin $\neq 0$, the matrix elements $\mathbf{U}^{(\varsigma,s)}_{l, k, m ; l^{\prime\prime\prime}, k^{\prime\prime\prime}, m^{\prime\prime\prime}} (g) $, are given by \eqref{smatrixelements} in the orthogonal basis
of the (Segal-)Bargmann-Fock Hilbert space $\mathcal{F}^{(\vsi,s)}$ composed by the$(2s+1)$-vector-valued analytic 
$\mathrm{F}^{(\vsi,s)}_{l,k,J,M}(\pz) $ \eqref{Enus0}.  

When we work with the elements $g^{}_{(b=0)}$, we get that the matrix elements, 
 $\textbf{U}^{(\varsigma,s)}_{l, k, J, M ; l, k^{\prime}, J^\prime, M^{\prime}} \big(g^{}_{(b=0)}\big)$, which are now given \eqref{kabab1}.

The character of the representation $U^{(\vsi,s)}$ is given by $\mbox{Tr}\,\textbf{U}^{(\varsigma,s)}(g^{}_d) = \sum_{l, k, J, M} \textbf{U}^{(\varsigma,s)}_{l, k, J, M ; l, k, J, M}(g^{}_d)$ \eqref{characters} with the matrix elements $\textbf{U}^{(\varsigma,s)}_{l, k, J, M ; l, k, J, M}(g^{}_d)$ expressed by \eqref{smatrixelementgd}.  
\end{enumerate}

\subsection*{Acknowledgments}
Mariano A. del Olmo is supported by  MCIN with funding from the European Union Next-GenerationEU (PRTRC17.I1), and also by PID2020-113406GB-I00  and PID2023-149560NB-C21 financed by MICIU/AEI/10.13039/501100011033 of Spain. Hamed Pejhan is supported by the Bulgarian Ministry of Education and Science, Scientific Programme ``Enhancing the Research Capacity in Mathematical Sciences (PIKOM)", No. DO1-67/05.05.2022. J.P. Gazeau would like to thank the University of Valladolid for its hospitality. The authors express their gratitude to Mohammad Enayati for his invaluable contributions during the initial stages of this manuscript. This article/publication is based upon work from COST Action CaLISTA CA21109 supported by COST (European Cooperation in Science and Technology).

\appendix

\setcounter{equation}{0} \section{Complex quaternions}\label{App. quaternions}

The set of complex quaternions $\mathbb{H}_{\mathbb{C}} \sim \C^4$ is defined as:
\begin{align}
\mathbb{H}_{\mathbb{C}} = \left\{ z= \underbrace{(z^4,\pmb{z})}_{\mbox{\small{scalar-vector notations}}} \equiv \quad \underbrace{z^4 + z^1\pmb{e}_1 + z^2\pmb{e}_2 + z^3\pmb{e}_3}_{\mbox{\small{Euclidean metric notations}}} \;\;;\;\; z^1,z^2,z^3,z^4 \in \C \right\}\,,
\end{align}
where $\pmb{e}_{i=1,2,3}$ obeys the following quaternionic algebra:
\begin{equation}\begin{array}{lll}\label{quatal}
\pmb{e}_i\pmb{e}_j &= \epsilon_{ij}^{\,\,\,k} \pmb{e}_k\,,\\[0.3cm]
\pmb{e}_i^2 &= -(1,\pmb{0}) \equiv -1 \quad (\mbox{by abuse of notations, we define $(1,\pmb{0}) \equiv 1$})\,,
\end{array}\end{equation}
with $\epsilon_{ij}^{\,\,\,k}$ ($i,j,k = 1,2,3$) representing the three-dimensional totally antisymmetric Levi-Civita symbol. 

The multiplication rule for two complex quaternions is as follows:
\begin{equation}\label{zzp}
zz^{\prime} = (z^4,\pmb{z})({z^{\prime}}^4,\pmb{z}^{\prime}) = \left(z^4{z^{\prime}}^4 -\pmb{z}\cdot\pmb{z}^{\prime}, z^4\pmb{z}^{\prime} + {z^{\prime}}^4\pmb{z} + \pmb{z}\times\pmb{z}^{\prime}\right)\,,
\end{equation}
where $\pmb{z}\cdot\pmb{z}^{\prime}$ and $\pmb{z}\times\pmb{z}^{\prime}$ denote, respectively, the analytic continuations of the Euclidean inner product and the cross product in $\mathbb{R}^3$.

For a given complex quaternion $z$, we can define:
\begin{equation}\label{conjugate}\begin{array}{lllll}
\hskip-3cm \bullet \quad &\text{complex conjugate:} \qquad &\overline z := \left(\overline{z^4},\overline{\pmb{z}} \right),\qquad  \text{with the property} \;\;\overline{{zz^\prime}} = \overline z\,\overline{{z^\prime}}
\\[0.3cm]
\hskip-3cm \bullet\quad &\text{quaternionic conjugate:} \qquad &\widetilde{z} := (z^4, -\pmb{z}),\qquad  \text{with the property} \;\;\widetilde{{zz^\prime}} = \widetilde{{z^\prime}} \widetilde z
\\[0.3cm]
\hskip-3cm \bullet\quad &\text{adjoint:} \qquad & z^{\ast} := \overline{\widetilde z} = \widetilde{\overline z},\hskip0.25cm\qquad  \text{with the property} \;\;({zz^\prime})^\ast = {z^{\prime\ast}} z^\ast
\\[0.3cm]
\hskip-3cm \bullet\quad &\text{determinant:} \qquad &\det z = \det \widetilde z = z\widetilde z = \widetilde z z = (z^1)^2 + (z^2)^2 + (z^3)^2 + (z^4)^2\,.
\end{array}\end{equation}
From the latter identity, one promptly obtains the expression for the inverse $z^{-1}$ of $z$ (assuming $\det z \neq 0$):
\begin{equation} \label{invcq}
z^{-1} = \frac{\widetilde z}{\det z}\,, \quad \mbox{with the property} \;\; \big( z z^\prime\big)^{-1} = {z^\prime}^{-1} z^{-1}\,.
\end{equation}

Let $M=\begin{pmatrix} a & b \\ c & d \end{pmatrix}$ be a generic $2\times 2$ matrix with components $a,b,c,d \in \H_{\C}$. 
The following four equivalent expressions, properly extended in case $a,b,c$, and/or $d$ are non-invertable, give the determinant of $M$ in terms of the determinant of its quaternionic components:
\begin{align} \label{det}
\det M = \big(\det a\big) \det (d-ca^{-1}b) = \big(\det b\big) \det (c-db^{-1}a) = \big(\det c\big) \det (b-ac^{-1}d) = \big(\det d\big) \det (a-bd^{-1}c)\,.
\end{align}

The algebra of complex quaternions $\H_{\C}$ is isomorphic to that of $2\times2$-complex matrices (denoted by $\mathcal{M}_2(\C)$). This isomorphism can be seen through the correspondences $(1,\pmb{0}) \equiv 1 \mapsto \bu_2$, $\pmb{e}_1\mapsto \ii \sigma_1$, $\pmb{e}_2\mapsto -\ii \sigma_2$, $\pmb{e}_3\mapsto \ii \sigma_3$, where $\sigma_{i=1,2,3}$ stands for the ordinary Pauli matrices, as:
\begin{equation} \label{cqumat}
\H_{\C}\ni z \;\mapsto\; Z(z) =
\begin{pmatrix} z^4 + \ii z^3 & \ii z^1 -z^2 \\\ii z^1+z^2 & z^4 - \ii z^3 \end{pmatrix} \equiv Z \in\mathcal{M}_2(\C)\,, \quad \det Z = \det z\,.
\end{equation}
It is worth noting that the multiplicative subgroup of real quaternions $\H \sim \R^4$ with unit norm is isomorphic to SU$(2) \sim \mathbb{S}^3$. For further elaboration on this topic, refer to Ref. \cite{Gazeau2022}.

\setcounter{equation}{0} \section{Matrix elements of the SU$(2)$ UIRs and their holomorphic extension}\label{App. SU(2)}

\subsection{Matrix elements of the SU$(2)$ UIRs}
Let $\xi = \begin{pmatrix} \xi^4 + \ii \xi^3 & - \xi^2 + \ii \xi^1 \\ \xi^2 + \ii \xi^1 & \xi^4 - \ii \xi^3 \end{pmatrix} \in \mathrm{SU}(2)$, that is, $\xi \equiv \big(\xi^4, \pmb{\xi}= (\xi^{i=1,2,3})\big) \in \H \sim \R^4$ with $\det \xi = 1$. Referring to the Talman notations \cite{talman68}, the matrix elements of the SU$(2)$ UIRs, $D^j$, are given by:
\begin{equation}\label{matelsu2}\begin{array}{lll}
D^j_{m_1 m_2} (\xi) &=& (-1)^{m_1-m_2} \sqrt{(j+m_1)! \; (j-m_1)! \; (j+m_2)! \; (j-m_2)!} \\[0.3cm]
& &\ds\quad\times \sum_{t} \frac{(\xi^4 + \ii \xi^3)^{j-m_2 - t}}{(j - m_2 -t)!} \, \frac{(\xi^4 - \ii\xi^3)^{j + m_1 - t}}{(j + m_1 -t)!}\,\frac{(- \xi^2 + \ii \xi^1)^{t+m_2 - m_1}}{(t + m_2 - m_1)!} \frac{(\xi^2 + \ii \xi^1)^{t}}{t!} \,,
\end{array}\end{equation}
where $j\in\mathbb{N}/2$, $-j \leqslant m_1,m_2 \leqslant j$, and $t$ (the values summed over) are such that the factorial functions remain non-negative, naturally compatible with $m_1-m_2\leqslant t \leqslant j+m_1$ and $0\leqslant t \leqslant j-m_2$.

There are two equivalent formulas for the reduction of the tensor product of two SU$(2)$ UIRs, containing the so-called $3-j$ symbols ($\propto$ Clebsch-Gordan coefficients):
\begin{equation}\label{CGdev2} \begin{array}{lll}
D^j_{m_1 m_2}(\xi)\,D^{j^\prime}_{m^\prime_1 m^\prime_2} (\xi) & =& \ds
\sum_{j^{\prime\prime} m^{\prime\prime}_1 m^{\prime\prime}_2}(2j^{\prime\prime} + 1)
\begin{pmatrix} j & j^\prime & j^{\prime\prime} \\ m_1 & m^\prime_1 & m^{\prime\prime}_1 \end{pmatrix}
\begin{pmatrix} j & j^\prime & j^{\prime\prime} \\ m_2 & m^\prime_2 & m^{\prime\prime}_2 \end{pmatrix}\, \overline{D^{j^{\prime\prime}}_{m^{\prime\prime}_1 m^{\prime\prime}_2} (\xi)}\,,\\[0.4cm]
&=&\ds \sum_{j^{\prime\prime} m^{\prime\prime}_1 m^{\prime\prime}_2}(2j^{\prime\prime} + 1) (-1)^{m^{\prime\prime}_1 - m^{\prime\prime}_2}\,\begin{pmatrix} j & j^\prime & j^{\prime\prime} \\ m_1 & m^\prime_1 & - m^{\prime\prime}_1 \end{pmatrix}
\begin{pmatrix} j & j^\prime & j^{\prime\prime} \\ m_2 & m^\prime_2 & - m^{\prime\prime}_2 \end{pmatrix} \, D^{j^{\prime\prime}}_{m^{\prime\prime}_1 m^{\prime\prime}_2} (\xi)\,.
\end{array}\end{equation}
Moreover, for any three representations there exists the following relation: 
\begin{equation}
\int_{SU(2)} D^j_{m_1 m_2} (\xi)\,D^{j^\prime}_{m^\prime_1 m^\prime_2} (\xi)\, D^{j^{\prime\prime}}_{m^{\prime\prime}_1 m^{\prime\prime}_2} (\xi)\, \ud\mu(\xi) = 2 \pi^2 \,
\begin{pmatrix} j & j^\prime & j^{\prime\prime} \\ m_1 & m^\prime_1 & m^{\prime\prime}_1 \end{pmatrix}
\begin{pmatrix} j & j^\prime & j^{\prime\prime} \\ m_2 & m^\prime_2 & m^{\prime\prime}_2 \end{pmatrix}.
\end{equation}
 On the other hand, the following expression for the $3-j$ symbols of Wigner, in the convention that they are all real, holds:
\begin{align} \label{3japc}
\nonumber \begin{pmatrix} j & j^\prime & j^{\prime\prime} \\ m & m^\prime & m^{\prime\prime} \end{pmatrix} &= (-1)^{j-j^\prime-m^{\prime\prime}} \sqrt{\frac{(j +j^\prime-j^{\prime\prime})!\; (j -j^\prime+j^{\prime\prime})!\; (-j +j^\prime +j^{\prime\prime})!}{(j +j^\prime +j^{\prime\prime} +1)!}}\\[0.3cm]
\nonumber&\quad\times \sqrt{(j+m)!\; (j-m)!\; (j^\prime+m^\prime)!\; (j^\prime-m^\prime)!\; (j^{\prime\prime}+m^{\prime\prime})!\; (j^{\prime\prime}-m^{\prime\prime})!} \\[0.3cm]
\nonumber&\quad\times \sum_s (-1)^s  \frac{1}{s!\; (j^\prime+m^\prime -s)!\; (j-m-s)!\; (j^{\prime\prime}-j^\prime+m +s)!}\\
&\mbox{\hskip 2cm}\times \frac{1}{(j^{\prime\prime} -j -m^\prime +s)!\; (j + j^\prime -j^{\prime\prime}-s)!} \nonumber\\[0.3cm]
&\equiv (-1)^{j-j^{\prime}-m^{\prime\prime}}(2j^{\prime\prime}+1)^{-1/2}\,\left(j m j^{\prime} m^{\prime}|j j^{\prime} j^{\prime\prime} -m^{\prime\prime}\right)\,,
\end{align}
with the constraints:
\begin{equation}
m + m^{\prime} + m^{\prime\prime}=0\,, \qquad \vert j-j^{\prime}\vert \leqslant j^{\prime\prime}\leqslant j+ j^{\prime}\,, 
\qquad j +j^{\prime} - j^{\prime\prime}\in \Z\,.
\end{equation}
Note that: (i) The values of $s$ (over which the summation is performed) ensure that the factorial functions remain non-negative. (ii) The expression $\left(j m j^{\prime} m^{\prime}|j j^{\prime} j^{\prime\prime} -m^{\prime\prime}\right)$ refers to the so-called \textit{vector-coupling} or \textit{Clebsch-Gordan} coefficients \cite{edmonds96}. (iii) The above expression in the particular case $m=m^\prime=m^{\prime\prime}=0$ turns into:
\begin{equation}\label{part3j}
\begin{split}
\begin{pmatrix} l_1 & l_2 & l_3 \\ 0 & 0 & 0 \end{pmatrix} = (-1)^{J/2} \left[\frac{(J-2l_1)!(J-2l_2)!(J-2l_3)!}{(J+1)!}\right]^{1/2} \frac{\left(\frac{J}{2}\right)!}{\left(\frac{J}{2}-l_1\right)!\left(\frac{J}{2}-l_2\right)!\left(\frac{J}{2}-l_3\right)!}\,,
\end{split}
\end{equation}
where $J\equiv l_1+l_2+l_3$. This expression vanishes if $J$ is odd.

\subsection{Holomorphic extension}
The holomorphic extension of the matrix elements \eqref{matelsu2} to $\H_{\C} \sim \C^4$ (see the notations, introduced in appendix A) is as follows:
\begin{equation}\begin{array}{lll} \label{matelHC} 
D^j_{m_1 m_2} (z) &=&\ds (-1)^{m_1-m_2} \sqrt{(j+m_1)! \; (j-m_1)! \; (j+m_2)! \; (j-m_2)!} \\[0.3cm]
&&\quad\ds \times \sum_{t} \frac{(z^4 + \ii z^3)^{j-m_2 - t}}{(j - m_2 -t)!} \, \frac{(z^4 - \ii z^3)^{j + m_1 - t}}{(j + m_1 -t)!}\,\frac{(- z^2 + \ii z^1)^{t+m_2 - m_1}}{(t + m_2 - m_1)!} \frac{(z^2+ \ii z^1)^{t}}{t!} \,,
\end{array}\end{equation}
where $z = \big(z^4,\pmb{z}=(z^{i=1,2,3})\big)\in \mathbb{H}_{\C} $.
The matrix elements $D^j_{m_1 m_2} (z)$ are harmonic polynomials, i.e., 
\begin{equation}\label{harmpol4}
\sum_{a=1}^4\frac{\partial^2}{\partial (z^a)^2} D^j_{m_1 m_2} (z) = 0\,.
\end{equation}
In this context, three key expansion identities are ($z,z^{\prime} \in \H_{\C}$): 
\begin{align}
\label{addexp1}
&\sigma_{m_1, m_2}^j\, D^j_{m_1 m_2} (z + z^{\prime}) = \sum_{j^\prime m^\prime_1 m^\prime_2} \sigma_{m_1-m^\prime_1\, m_2 -m^\prime_2}^{j-j^\prime} \,D_{m_1-m^\prime_1\, m_2 -m^\prime_2}^{j-j^\prime}(z) \,\sigma_{m^\prime_1 ,m^\prime_2}^{j'} \,D_{m^\prime_1 m^\prime_2}^{j^\prime} (z^{\prime})\,,
\\[0.25cm]
\label{addexp2}
& \left(\sigma_{m_1, m_2}^j\right)^{-1}\, \big[\det(z+z^{\prime})\big]^{-1} \,D^j_{m_1 m_2} \left((z + z^{\prime})^{-1}\right) = \nonumber \\
&\quad\sum_{j^\prime m^\prime_1 m^\prime_2} (-1)^{2j^\prime} \sigma_{m^\prime_1, m^\prime_2}^{j^\prime} \,D_{m^\prime_1 m^\prime_2}^{j^\prime} (z)\,\left(\sigma_{m_1 +m^\prime_2\, m_2 +m^\prime_1}^{j +j^\prime}\right)^{-1} \,(\det z^{\prime})^{-1}\,D_{m_1 +m^\prime_2\, m_2 +m^\prime_1}^{j +j^\prime}({z^{\prime}}^{-1}) \,, \quad \det z < \det z^{\prime} \,, \\[0.25cm]
\label{multexp}  
&D^j_{m_1 m_2} (z z^{\prime}) = \sum_{m^\prime}D^j_{m_1 m^\prime} (z)\,D^j_{m^\prime m_2} ( z^{\prime}) \, ,
\end{align}
where:
\begin{align} \label{sjm1}
\sigma_{m_1 ,m_2}^j = \sigma_{m_1 }^j \, \sigma_{ m_2}^j\,, \quad \mbox{with} \quad \sigma_m^j = \frac{1}{\sqrt{(j-m)!\,(j+m)!}} \,.
\end{align}
Note that: (i) The indices to be summed over are chosen from the allowed ranges corresponding to the indices of the SU$(2)$ representations. (ii) The first two expansions are addition theorems, while the latter arises directly from the group representation. Detailed explanations can be found in Ref. \cite{gazeau78}.

The connection between $\mathcal{Y}_{lm}(\pmb{z})$, the holomorphic extension of the solid spherical harmonics, and the matrix elements of the SU$(2)$ UIRs is established as \cite{hassan80}:
\begin{equation} \label{YlDl2}
\big(\sigma^l_m\big)^{-1}\,\mathcal{Y}_{lm}(\pmb{z}) = 2^{-l}\left(\frac{2l+1}{4\pi}\right)^{1/2} \sum_{\stackrel{m_1 m_2}{m_2-m_1=m}}(-1)^{m_1}\; \big(\sigma^{l/2}_{m_1,m_2}\big)^{-1}\; D^{l/2}_{m_1m_2}[(z^4,\pmb{z})]\,,
\end{equation}
where the value of $z^4$ is arbitrary. Conversely, we have:
\begin{equation} \label{DlYl2}
\begin{split}
{D}^{l/2}_{m_1m_2}[(0,\pmb{z})] \underbrace{\equiv \;\; {D}^{l/2}_{m_1m_2}(\pmb{z}) \;\;}_{\mbox{by abuse of notations!}} &= (4\pi)^{1/2}\sum_{\stackrel{l^{\prime}}{l-l^{\prime} = \mathrm{even}}}(-1)^{l^{\prime}+m_2}\; 2^{l^{\prime}}\; (2l^{\prime}+1)^{1/2} \left(\frac{(l-l^{\prime})!}{(l+l^{\prime}+1)!}\right)^{1/2} \frac{\left[\frac{1}{2}(l+l^{\prime})\right]!}{\left[\frac{1}{2}(l-l^{\prime})\right]!}\\
&\qquad\qquad\quad \times \begin{pmatrix} \frac{l}{2} & \frac{l}{2}& l^{\prime} \\ m_1 & -m_2 & m^{\prime} \end{pmatrix} (\pz\cdot\pz)^{\frac{l-l^{\prime}}{2}}\mathcal{Y}_{l^{\prime}m^{\prime}}(\pmb{z})\,.
\end{split}
\end{equation}

\section{A useful expansion formula}\label{App. expansion}
We begin by recalling the generating function for Gegenbauer polynomials \cite{hua63}:
\begin{equation} \label{expgeg1}
(1 + u^2 - 2 u t)^{-\lambda} = \sum_{l=0}^{\infty} u^l \, C_l^{\lambda}(t)\,, \quad \vert u \vert <1\,, \quad \lambda \neq 0\,.
\end{equation}
Note that, for negative integer values of $\lambda$, the left-hand side of the above identity becomes a finite sum:
\begin{align}\label{footnote negative lambda0}
(1 + u^2 - 2 u t)^{-\lambda} = \sum_{l=0}^{2|\lambda|} u^l \, C_l^{\lambda}(t)\,.
\end{align}

Utilizing the first identity, we expand $\left[\det\left(1 + \pz\,\overline{\pz^{\prime}}\right)\right]^{-\lambda}$:
\begin{align}\label{opop}
\left[\det\left( 1 + \pz\,\overline{\pz^{\prime}}\right)\right]^{-\lambda}&=  \left[1 - 2 \pz\cdot\overline{\pz^{\prime}} + \left(\pz\cdot\pz\right)\overline{\left(\pz^{\prime}\cdot\pz^{\prime}\right)}\right]^{-\lambda} \nonumber\\
&= \left[1 - 2 \left[\left(\pz\cdot\pz\right)\overline{\left(\pz^{\prime}\cdot\pz^{\prime}\right)}\right]^{1/2} \frac{\pz\cdot\overline{\pz^{\prime}}}{\left[\left(\pz\cdot\pz\right)\overline{\left(\pz^{\prime}\cdot\pz^{\prime}\right)}\right]^{1/2}}+ \left(\pz\cdot\pz\right)\overline{\left(\pz^{\prime}\cdot\pz^{\prime}\right)}\right]^{-\lambda} \nonumber\\
&= \sum_{l=0}^{\infty} \left[\left(\pz\cdot\pz\right)\overline{\left(\pz^{\prime}\cdot\pz^{\prime}\right)}\right]^{l/2} \, C_l^{\lambda}\left(\frac{\pz\cdot\overline{\pz^{\prime}}}
{\left[\left(\pz\cdot\pz\right)\overline{\left(\pz^{\prime}\cdot\pz^{\prime}\right)}\right]^{1/2}}\right)\,.
\end{align}
The square root of the holomorphic quadratic $\left(\pz\cdot\pz\right)$ mentioned above is understood to lie on the first Riemann sheet.

Next, we employ the expansion of Gegenbauer polynomials \cite{hua63}, expressed in relation to the holomorphic extensions of the solid spherical harmonics (for the latter, see Eq. \eqref{solspharmp}):
\begin{align}\label{gegen1}
& C_l^{\lambda}(t) = \frac{\sqrt{\pi}}{\Gamma(\lambda)\; \Gamma\left(\lambda-\frac{1}{2}\right)}\,\sum_{k = 0}^{\lfloor \frac{l}{2}\rfloor} d_k\,  C_{l-2k}^{1/2} (t)\,, \qquad 
d_k = \frac{\left(l-2k + \frac{1}{2}\right)\, \Gamma\left(k+\lambda - \frac{1}{2}\right)\, \Gamma(\lambda + l - k)}{k! \,\Gamma\left(l-k + \frac{3}{2}\right)} \,,\\[0.25cm]
\label{geghyp1} &\left[\left(\pz\cdot\pz\right)\overline{\left(\pz^{\prime}\cdot\pz^{\prime}\right)}\right]^{l/2} \, C_l^{1/2}
\left(\frac{\pz\cdot\overline{\pz^{\prime}}}{\left[\left(\pz\cdot\pz\right)\overline{\left(\pz^{\prime}\cdot\pz^{\prime}\right)}\right]^{1/2}}\right) = \frac{4\pi}{2l + 1} \sum_{m=-l}^l \mathcal{Y}_{lm}(\pz)\,\overline{\mathcal{Y}_{lm}(\pz^{\prime})}\,.
\end{align}
Having in mind the functional relation of the Gamma functions,  $\Gamma(x+1)=x\, \Gamma(x)$, we may also introduce the following expansions:
\begin{equation}
\Gamma\big(k+\lambda-{1}/{2}\big) =
(\lambda -1/2)_k \,\Gamma\big(\lambda-{1}/{2}\big)\,,
\quad
\Gamma\big(\lambda + l-k\big) = (\lambda)_{l-k} \,\Gamma(\lambda)\,,
\end{equation}
where $(x)_k$ is the Pochhammer symbol: 
\begin{equation}\label{pochhammer}
(x)_k:=x(x-1)\dots (x+k-1)\,,\qquad  (x)_0:=1\,. 
\end{equation}
Then, we have also an alternative form of Eq.~\eqref{gegen1} as:
\begin{equation}\label{234} 
C_l^{\lambda}(t) =\ds {\sqrt{\pi}}\,\sum_{k = 0}^{\lfloor \frac{l}{2}\rfloor} d^\prime_k\, C_{l-2k}^{1/2} (t)\,,
\end{equation}
with:
\begin{equation}\label{235}
d^\prime_k =\ds \frac{\left(l-2k + \frac{1}{2}\right)\,(\lambda -1/2)_k \, (\lambda)_{l-k} }{k! \,\Gamma\left(l-k + \frac{3}{2}\right)}\, 
\end{equation}
Unlike the previous identity \eqref{gegen1}, the omission of terms such as $\Gamma(\lambda-1/2)$ and $\Gamma(\lambda)$ in the latter permits consideration for negative values of $\lambda$ as well.

Now, as we combine Eqs.~\eqref{opop}, \eqref{gegen1}, and \eqref{geghyp1}, while invoking the \emph{Legendre duplication formula}:
\begin{align}
\Gamma(x) \; \Gamma(x+1/2) = 2^{1-2x} \sqrt{\pi} \; \Gamma(2x)\,,
\end{align}
the subsequent expansion (applicable for $\lambda \neq 0$) emerges prominently:
\begin{align} 
\label{expdetlambd'} \left[\det\left( 1 + \pz\,\overline{\pz^{\prime}}\right)\right]^{-\lambda} &= \sum_{l=0}^{\infty}\sum_{k = 0}^{\lfloor \frac{l}{2}\rfloor} \sum_{m=2k-l}^{l-2k}\,a_{\lambda,l,k}\,(\pz\cdot\pz)^{k}\, \mathcal{Y}_{l-2k,m}(\pz)\,\overline{(\pz^{\prime}\cdot\pz^{\prime})}^{k}\,\overline{\mathcal{Y}_{l-2k,m}(\pz^{\prime})}\,,\\[0.3cm]
\label{normlk} a_{\lambda,l,k} &= \frac{2^{2\lambda-1} \,\pi \,\Gamma\left(k+\lambda - \frac{1}{2}\right) \Gamma(\lambda + l - k)}{\Gamma(2\lambda-1)\,k! \,\Gamma\left(l-k + \frac{3}{2}\right)}\,.
\end{align}


\end{document}